\documentclass[twocolumn]{aastex701}

\usepackage{fontawesome}

\begin{document}

\title{Universal Fitting Formulae for the Peak Concentration of Dark Matter Halos}

\author[orcid=0000-0002-5358-3305,sname='Wang']{Dao-zhou Wang}
\affiliation{School of Physics and Astronomy, Sun Yat-sen University, DaXue Road 2, 519082 Zhuhai, China}
\email{wangdaozhou12@163.com}  

\author[orcid=0000-0003-2204-2474, sname='Lin']{Weipeng Lin} 
\affiliation{School of Physics and Astronomy, Sun Yat-sen University, DaXue Road 2, 519082 Zhuhai, China}
\affiliation{CSST Science Center for the Guangdong-Hongkong-Macau Greater Bay Area, DaXue Road 2, 519082 Zhuhai, China}
\email[show]{linweip5@mail.sysu.edu.cn}

\author[orcid=0009-0001-6592-6467,sname='Luan']{Tian-Cheng Luan} 
\affiliation{School of Physics and Astronomy, Sun Yat-sen University, DaXue Road 2, 519082 Zhuhai, China}
\email{dingdluan@outlook.com}

\begin{abstract}
The prediction of the structural properties of dark matter halos is crucial for studies in modern cosmology and galaxy formation. Utilizing a comprehensive suite of $N$-body simulations spanning diverse cosmologies and box sizes, we derive a universal halo concentration prescription aligned with the excursion set theory framework. The halo peak-height parameter is revised to incorporate the linear growth factor at the halo formation epoch, enhancing its physical relevance to assembly history tracking. For fixed halo mass, we fit the distributions of the halo concentration and revised peak height to lognormal functions to extract their peak values. We find that these peaks follow a universal, tight relation invariant to redshift, box size, initial power spectrum, and cosmology, exhibiting remarkably small scatter. 
In particular, the relation flattens systematically with increasing revised peak height, 
approaching an asymptotic value of $\sim$2.54, consistent with previous reports of a concentration floor. The fitting formula consistently describes both the mass$-$concentration and peak height$-$concentration relations, easily quantifying how these relations depend on redshift and cosmological parameters. This universal framework enables robust prediction of the most probable (peak) halo concentration using our fitting formula or theoretical/semianalytical mass assembly histories. A software package for calculating peak concentrations is publicly available online.
\end{abstract}


\section{Introduction} \label{sec:intro}
It is widely accepted that the structure of dark matter halos can be well described by a two-parameter density profile proposed by Navarro, Frenk, and White (\citealp{1996ApJ...462..563N, 1997ApJ...490..493N}, hereafter NFW).
Numerous studies have sought to quantify halo concentration and its dependence on dynamical state, environment, halo mass, redshift, formation history, density fluctuation power spectrum, and cosmological model (e.g., \citealp{2000ApJ...535...30J}; \citealp{2001MNRAS.321..559B}; \citealp{2001ApJ...554..114E}; \citealp{2002ApJ...568...52W}; \citealp{2003ApJ...597L...9Z}; \citealp{2005MNRAS.357...82R};  \citealp{2007MNRAS.378...55M}; \citealp{2007MNRAS.381.1450N}; \citealp{2008MNRAS.390L..64D}; \citealp{2008MNRAS.387..536G}; \citealp{2012MNRAS.427.1322L, 2014MNRAS.441..378L}; \citealp{2014MNRAS.441.3359D};  \citealp{2015ApJ...799..108D}; \citealp{2019ApJ...871..168D}).

Significant efforts have also been made to develop prescriptions for describing or predicting halo concentration (e.g., \citealp{1997ApJ...490..493N}; \citealp{2001MNRAS.321..559B}; \citealp{2003MNRAS.339...12Z, 2009ApJ...707..354Z}; \citealp{2008MNRAS.391.1940M}; \citealp{2011ApJ...740..102K}; \citealp{2012MNRAS.423.3018P}; \citealp{2013ApJ...766...32B}; \citealp{2014MNRAS.441..378L, 2016MNRAS.460.1214L}; \citealp{2015ApJ...799..108D}; \citealp{2019ApJ...871..168D}). 
Specifically, \cite{1997ApJ...490..493N} proposed that concentration depends on the time at which a fixed fraction of the halo mass is assembled. Although their original model was found to inaccurately predict concentration evolution (see \citealp{2001MNRAS.321..559B}), the core concept$-$linking concentration to halo mass assembly history (MAH)$-$was widely validated in subsequent studies (e.g., \citealp{2002ApJ...568...52W}; \citealp{2003ApJ...597L...9Z, 2009ApJ...707..354Z}; \citealp{2012MNRAS.427.1322L, 2014MNRAS.441..378L}). Notably, \cite{2016MNRAS.460.1214L} developed an analytical model that can accurately predict the concentrations of individual halos in both $\Lambda$CDM and $\Lambda$WDM models, given detailed information on the ‘collapsed mass history’ (for all progenitors).

Along these lines, one of the useful methods utilizes the so-called halo peak height ($\nu$, see Equation (\ref{nu})), a concept rooted in excursion set theory (\citealp{1986ApJ...304...15B}; \citealp{1991ApJ...379..440B}; \citealp{1993MNRAS.262..627L}; \citealp{2000MNRAS.319..168C, 2008MNRAS.383..546C}) that links halo formation to both the spherical-collapse threshold and the rms density fluctuation.

Simulation results show that, while the dependence of halo concentration on cosmology and redshift can be effectively parameterized by $\nu$, the universality of the $c$--$\nu$ relation remains limited across different cosmological models and redshift regimes (e.g., \citealp{2013ApJ...766...32B}; \citealp{2014MNRAS.441..378L}; \citealp{2014MNRAS.441.3359D}). This cosmology and redshift dependence indicates missing parameters in current models, yet the quest for universality persists.

In merger trees derived from the extended Press-Schechter (EPS) formalism, where only halo mass is explicitly incorporated, the halo formation-redshift distribution can be computed using the conditional probability function (e.g., \citealp{1993MNRAS.262..627L}). Simulations show that this distribution typically follows a near-lognormal profile (e.g., \citealp{2003MNRAS.344.1327L}), characterized by a peak probability mode, or peak value (hereafter PV), that represents the most probable formation redshift. Similarly, concentrations in matching mass bins also exhibit lognormal distributions (e.g., \citealp{2000ApJ...535...30J}; \citealp{2001MNRAS.321..559B}; \citealp{2024MNRAS.52710760W}). We thus propose that these shared features may create a universal correlation between concentration and formation time PVs.

Inspired by the concept above, we modify the definition of halo peak height by implicitly incorporating the halo formation time $z_\mathrm{f}$ (defined as the redshift at which the halo reaches half of its final mass at redshift $z$) and the halo mass at that redshift, aiming to better capture the $c$--$\nu$ relation.
By analyzing the results of a suite of cosmological simulations and performing fits across all halos at multiple redshifts, we successfully derive a simple, universal prescription for the PV of the concentration, expressed in terms of the PV of the revised peak height\footnote{Here, we argue that for a lognormal distribution, the mode (peak value) is more representative than the median and mean, following the typical ordering: mode $<$ median $<$ mean.}. 
This framework enables prediction of the peak halo concentration as the MAH is known.  

The paper is organized as follows. In Section \ref{data_and_method}, we describe the simulations, data, and methodology. The results are presented in Section \ref{results}. 
The summary is given in Section \ref{summary}. Throughout this paper, the notation “log" denotes a base-10 logarithm.

\section{definition, data, and method} \label{data_and_method}

\subsection{Halo Concentration}

In this paper, a virial halo is defined as a region whose mean density within the halo radius $r_{h}$ is $\Delta_{h}$ times the mean cosmic density $\bar\rho(z)$ at redshift $z$. The corresponding halo mass is expressed as
\begin{equation}
M_{h} \equiv \frac{4\pi}{3}\Delta_{h}\bar{\rho}(z) r_{h}^{3}.
\end{equation}

The NFW profile is adopted to model the structure of such virial halos.
The ratio of the halo virial radius to the radius (generally denoted as $r_{-2}$ or $r_{s}$) at which the logarithmic slope is -2 defines the concentration parameter: 
\begin{equation}
   c_{\Delta}=r_{\Delta}/r_{s},
\end{equation}
which characterizes the shape of the density profile.
In some previous studies, $\Delta_{h}$ = 200 times the cosmic critical density has been adopted to identify halos, with the corresponding mass, radius, and concentration denoted as $M_{200c}$, $r_{200c}$ and $c_{200c}$, respectively. 
In this work, we primarily adopt the virial overdensity $\Delta_{h} = \Delta_{\mathrm{vir}}$, for which we use the fitting formulae provided by \citet{1998ApJ...495...80B}, although results for $c_{200c}$ are also provided.

\subsection{Numerical Simulations} \label{data}

A collection of publicly accessible $N$-body simulations, complete with halo catalogs and merger trees, has been compiled from the TNG project (\citealp{2018MNRAS.480.5113M}; \citealp{2018MNRAS.477.1206N}; \citealp{2018MNRAS.475..624N}; \citealp{2018MNRAS.475..648P}; \citealp{2018MNRAS.475..676S})\footnote{\url{https://www.tng-project.org}} and the MultiDark-Planck project (\citealp{2016MNRAS.457.4340K}).\footnote{\url{https://www.cosmosim.org}} 

To examine cosmological model dependence, a suite of simulations has been conducted by us, including standard CDM (SCDM) and open CDM (OCDM) models with parameter configurations adopted from \citet{1998ApJ...494L...5J, 2002ApJ...574..538J}, scale-free (SF) model (assuming a power-law initial power spectrum $P(k)\propto k^{n}$ with a spectral index $n=-2$) of an Einstein-de Sitter (EdS) universe, $w$CDM models, as well as a set of warm dark matter (WDM) models in the $\Lambda$WDM cosmology.
Note that the $w$CDM simulations adopt two values of the $w$ parameter that differ substantially, corresponding to the first ($w=-0.816$) and 37th ($w=-1.294$) models in the Coyote suite of universes (see \citealp{2009ApJ...705..156H}, \citealp{2010ApJ...715..104H} for details), while the 
$\Lambda$WDM simulations adopt equivalent thermal relic particle masses of 0.5, 0.8, 1.1, and 1.5 keV, respectively. Specifically, the simulations of the aforementioned non-$\Lambda$CDM models were carried out with the \texttt{GADGET-4} (\citealp{2021MNRAS.506.2871S}) and \texttt{SWIFT} codes (\citealp{2024MNRAS.530.2378S}).

The detailed settings for each simulation are summarized in Table \ref{simulation}.
These simulations span a broad range of box sizes (30$-$2500$\ h^{-1}\ \mathrm{Mpc}$) and mass resolutions (3.65$\times 10^{5}-$2.36$\times 10^{10}\ h^{-1}\ M_{\odot}$), enabling systematic investigations into the connections between halo structure, formation history, and diverse cosmological environments across mass scales.

The halo catalogs and merger trees for the TNG simulations are publicly available from the database described in \citet{2022MNRAS.509.3441A}. However, these catalogs provide only the concentration parameter $c_{200c}$. While $c_{200c}$
can be converted to the virial concentration $c_{\mathrm{vir}}$ via a standard method that assumes an NFW profile,  
we choose to perform independent profile fitting to treat the inner fitting radius consistently.
Previous studies have demonstrated that force and mass resolution in numerical simulations strongly affect the characterization of dark matter halo structures, and numerous convergence analyses have been widely performed (\citealp{2003MNRAS.338...14P, 2007MNRAS.381.1450N, 2010MNRAS.402...21N, 2019MNRAS.488.3663L}).
Specifically, \cite{2019MNRAS.488.3663L} concluded that the minimum radii for fitting an NFW profile should start from $r_{\rm conv}$, the so-called convergence radius, to minimize the impact of collisional relaxation. For most of the simulations utilized in this paper, we perform NFW profile fitting using 50 radial bins logarithmically spaced from $3 \epsilon$  to $r_{\Delta}$, as it was verified that $3\epsilon$ is larger than or comparable to the estimated $r_{\rm conv}$. However, for the TNG simulations, it is true that $3\epsilon < r_{\rm conv}$ and we perform the fitting with an inner radius of $r_{\rm conv}$.
The halo catalogs and merger trees for the MultiDark-Planck simulations were downloaded from the project's official website. For our simulations, following the MultiDark-Planck project, the halo catalogs (with concentration) and merger trees were generated using the phase-space halo finder Rockstar and the Consistent Trees code package (\citealp{2013ApJ...762..109B, 2013ApJ...763...18B}), respectively. 

To ensure robust density profile fitting and reliable estimation of the concentration parameter, we adopt a strict particle-number threshold and select only halos resolved by at least 2000 dark matter particles across all simulations, following a conservative criterion comparable to that adopted by \citet{2018ApJ...859...55C} for concentration measurements.
This threshold is more conservative than the resolution level discussed by \citet{2022MNRAS.509.3441A}, who found generally good numerical consistency in scaling parameters even at the low-mass end, where halos are resolved by $\sim 1500$ particles.
We employ a large suite of simulations, allowing halos of identical mass to be investigated at varying resolutions to assess the possible impact of resolution effects in overlapping mass ranges. Due to the diverse mass resolutions of our simulation suite, this universal particle threshold naturally translates into different minimum halo masses for each specific simulation. Consequently, the absolute minimum halo mass analyzed in this work is $\sim$7.3 $\times\ 10^8\ h^{-1}\ M_{\odot}$, which corresponds to the 2000-particle limit in our highest-resolution dataset, TNG50-dark.

\begin{table*}
\centering
\small 
\setlength{\tabcolsep}{0.5pt}
\renewcommand\arraystretch{1.0}
\caption{Overview of the Simulations Utilized in This Work}
\begin{tabular*}{\textwidth}{@{\extracolsep{\fill}} l c c c c c c c c c c c } 
\hline\hline
Simulation & $L_{\text {box }}$ & $N_{\text {DM }}$ & $m_{\text {DM }}$ & $\epsilon$ & $\Omega_{m}$ & $\Omega_{\Lambda}$ & $\sigma_{8}$ & $n_{s}$ & $h$ & $w$ or $m_\mathrm{WDM}$ & Code \\
& $\left(h^{-1}\ \mathrm{Mpc}\right)$ &  
& $\left(h^{-1}\ M_{\odot}\right)$ & $(h^{-1}\ \mathrm{kpc})$ &  &  &  &  &  & (keV) & \\
\hline 
TNG50-dark & $35$  &  $2160^{3}$  & 3.65$\ \times\ 10^{5}$ & 0.20 & 0.3089 & 0.6911 & 0.8159 & 0.9667 & 0.6774 & $\cdots$  & \texttt{AREPO}\\
TNG100-dark & $75$ & $1820^{3}$ & 6.0$\ \times\ 10^{6}$ & 0.50 & 0.3089 & 0.6911 & 0.8159 & 0.9667 & 0.6774 & $\cdots$  & \texttt{AREPO}\\
TNG300-dark & $205$ & $2500^{3}$ & 4.73$\ \times\ 10^{7}$ & 1.0 & 0.3089 & 0.6911 & 0.8159 & 0.9667 & 0.6774 & $\cdots$  & \texttt{AREPO}\\
VSMDPL & 160 & $3840^{3}$ & 6.20$\ \times\ 10^{6}$ & 2.0 & 0.307 & 0.693 & 0.8228 & 0.960 & 0.6777 & $\cdots$  & \texttt{GADGET-2}\\
SMDPL & 400 & $3840^{3}$ & 9.63$\ \times\ 10^{7}$ & 1.50 & 0.307 & 0.693 & 0.8228 & 0.960 & 0.6777 & $\cdots$  & \texttt{GADGET-2}\\
MDPL2 & 1000 & $3840^{3}$ & 1.51$\ \times\ 10^{9}$ & 5.0 & 0.307 & 0.693 & 0.8228 & 0.960 & 0.6777 & $\cdots$  & \texttt{GADGET-2}\\
BigMDPL & 2500 & $3840^{3}$ & 2.36$\ \times\ 10^{10}$ & 10.0 & 0.307 & 0.693 & 0.8228 & 0.960 & 0.6777 & $\cdots$  & \texttt{GADGET-2}\\
SCDM & 100 & $512^{3}$ & 2.07$\ \times\ 10^{9}$ & 4.88 & 1.0 & 0 & 0.55 & 1.0 & 0.50 & $\cdots$  & \texttt{GADGET-4}\\
OCDM & 100 & $512^{3}$ & 6.20$\ \times\ 10^{8}$ & 4.88 & 0.30 & 0 & 1.0 & 1.0 & 0.667 & $\cdots$  & \texttt{GADGET-4}\\
SF-n2 & 100 & $512^{3}$ & 2.07$\ \times\ 10^{9}$ & 4.88 & 1.0 & 0 & 1.0 & $-$2.0 &$ \cdots$ & $\cdots$  & \texttt{GADGET-4}\\
$\Lambda$WDM-L100-0.5 & 100 & $512^{3}$ & 6.43$\ \times\ 10^{8}$ & 4.88 & 0.31 & 0.69 & 1.0 & 1.0 & 0.6766 & 0.5 & \texttt{GADGET-4}\\
$\Lambda$WDM-L100-0.8 & 100 & $512^{3}$ & 6.43$\ \times\ 10^{8}$ & 4.88 & 0.31 & 0.69 & 1.0 & 1.0 & 0.6766 & 0.8 & \texttt{GADGET-4}\\
$\Lambda$WDM-L100-1.1 & 100 & $512^{3}$ & 6.43$\ \times\ 10^{8}$ & 4.88 & 0.31 & 0.69 & 1.0 & 1.0 & 0.6766 & 1.1 & \texttt{GADGET-4}\\
$\Lambda$WDM-L100-1.5 & 100 & $512^{3}$ & 6.43$\ \times\ 10^{8}$ & 4.88 & 0.31 & 0.69 & 1.0 & 1.0 & 0.6766 & 1.5 & \texttt{GADGET-4}\\
$\Lambda$WDM-L60-1.5 & 60 & $512^{3}$ & 1.39$\ \times\ 10^{8}$ & 2.93 & 0.31 & 0.69 & 1.0 & 1.0 & 0.6766 & 1.5 & \texttt{GADGET-4}\\
$\Lambda$WDM-L30-1.5 & 30 & $1024^{3}$ & 2.17$\ \times\ 10^{6}$ & 0.73 & 0.31 & 0.69 & 1.0 & 1.0 & 0.6766 & 1.5 & \texttt{GADGET-4}\\
$w$CDM1 & 100 & $512^{3}$ & 8.91$\ \times\ 10^{8}$ & 4.88 & 0.4308 & 0.5692 & 0.8161 & 0.9468 & 0.5977 & $-$0.816 & \texttt{SWIFT}\\
$w$CDM37 & 100 & $512^{3}$ & 5.79$\ \times\ 10^{8}$ & 4.88 & 0.28 & 0.72 & 0.9000 & 1.0233 & 0.7313 & $-$1.294 & \texttt{SWIFT}\\
\hline
\end{tabular*}

\vspace{1.5ex}
\begin{minipage}{\textwidth}
\textbf{Note.} The columns give the simulation identifier, the comoving box length $L_{\rm box}$, the number of dark matter particles $N_{\rm DM}$, the particle mass $m_{\rm DM}$, the Plummer-equivalent gravitational softening length $\epsilon$, the cosmological parameters $\Omega_m$, $\Omega_{\Lambda}$, $\sigma_8$, $n_s$, and $h$, the additional model parameter $w$ or $m_{\rm WDM}$ for $w$CDM and WDM cosmologies, respectively, and the simulation code.
\end{minipage}
\label{simulation}
\end{table*}

\subsection{Method} \label{method}

\begin{figure*}
\centering
 \includegraphics[width=16cm,height=7cm]{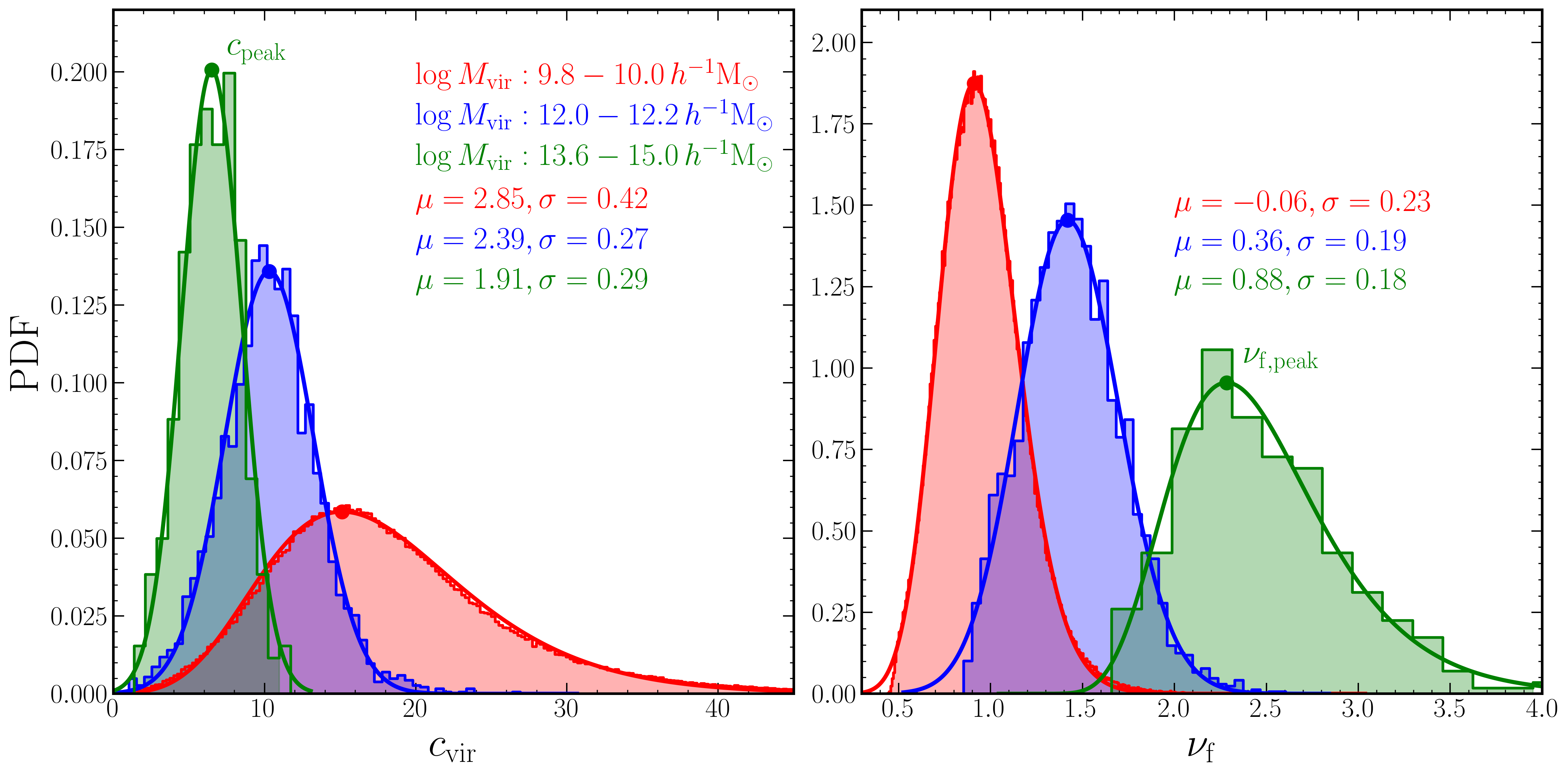}
\caption{Probability distribution functions of $c_\mathrm{vir}$ (left panel) and $\nu_\mathrm{f}$ (right panel) in three representative mass bins for the halos in the VSMDPL simulation. The distributions of both halo properties are fit with lognormal functions, shown as solid lines. The best-fit parameters and mass ranges used are shown in each panel. The peak values are defined as $c_\mathrm{peak}$ and $\nu_\mathrm{f,peak}$, respectively.
Both $\mu$ and $\sigma$ are computed in logarithmic space.
}
\label{para_define}
\end{figure*}

Halo formation processes are often characterized by the dimensionless peak height $\nu$ \citep{1986ApJ...304...15B, 1993MNRAS.262..627L, 
2013ApJ...766...32B, 2016MNRAS.457.4340K}, which is defined as
\begin{equation}
    \nu(M,z)=\frac{\delta_{sc}(z)}{\sigma(M)},
    \label{nu}
\end{equation}
where $\delta_{sc}(z)$($\equiv \delta_{sc}(0)/D(z)$) is the critical overdensity at redshift $z$ and $\sigma(M)$ is the rms fluctuation of the linear density field today ($z=0$) on a mass scale $M$. 
The present-day critical overdensity $\delta_{sc}(0)=1.686$ is adopted, ignoring minor cosmology- and redshift-dependent variations. 
Here, $D(z)$ ($\equiv D^+(z)/D^+(0)$) denotes the linear growth factor normalized to unity at $z=0$;
for detailed calculation procedures of $D^+(z)$, see \citet{2007ApJ...671.1160L}. 
Note that $\sigma(M)$ depends on the underlying cosmology through the linear fluctuation power spectrum, $P(k)$, which is computed using codes such as CAMB \citep{2000ApJ...538..473L}. The power spectrum is shaped by cosmological parameters, including the primordial fluctuation spectral index $n_s$, and the normalization $\sigma_8$.
The calculation of $\sigma(M)$ also depends on the window function adopted to smooth the density field.

The halo peak height $\nu$ has been used to formulate the empirical relation between peak height and the Einasto shape parameter $\alpha$ \citep{2008MNRAS.387..536G}, and to incorporate the joint dependence of halo concentration on redshift and mass (e.g., \citealp{2009ApJ...707..354Z}; \citealp{2012MNRAS.427.1322L, 2014MNRAS.441..378L}; \citealp{2015ApJ...799..108D}). Yet, this peak-height definition ignores the dependence on halo growth history, hindering the development of a universal model for the redshift evolution of halo concentration.
Thus, we revise the framework to incorporate halo MAH by defining a formation-time-dependent peak height as 
\begin{equation}
    \nu_\mathrm{f}(M,z,z_\mathrm{f})=\frac{\nu (M,z)}{D(z_\mathrm{f})}
    =\frac{\delta_{sc}(0)}{D(z_\mathrm{f}) D(z) \sigma(M)}.
    \label{nv_f}
\end{equation}
We refer to $\nu_\mathrm{f}$ as the formation peak height hereafter. 
Incorporating $D(z_\mathrm{f})$ is key to the revision of the peak height. 
The extra factor $D(z_\mathrm{f})^{-1}$ accounts for the tendency of earlier-forming halos to feature lower linear growth factors and generally denser inner profiles.
Thus, $\nu_\mathrm{f}$ can be viewed as an effective collapse rarity that incorporates both the identification epoch and characteristic assembly epoch of halos.
The use of these quantities facilitates universality, as their complex dependence on cosmological models can be naturally absorbed into the framework.

We next describe the treatment of $\sigma(M)$ in the $\Lambda$WDM models and introduce the half-mode mass used to characterize the suppression scale of the WDM power spectrum.
For the $\Lambda$WDM cosmology, we adopt a sharp-$k$ filter rather than a real-space top-hat filter, following \citet{2013MNRAS.433.1573S}. This choice is standard for WDM models and yields a reasonable halo mass function. The corresponding mass is assigned as $M = \frac{4\pi}{3}\,\bar{\rho}\,(c_{\rm sk}R)^3$,
where $\bar{\rho}$ denotes the cosmic mean matter density. We set $c_{\rm sk} = (9\pi/2)^{1/3} \approx 2.42$ following \citet{1993MNRAS.262..627L}, though somewhat larger calibrated values have also been adopted in other studies \citep[e.g.,][]{2013MNRAS.433.1573S}.
The half-mode mass, $M_{\mathrm{hm}}$, is defined following \citet{2001ApJ...559..516A} as the mass associated with the half-mode wavenumber $k_{\mathrm{hm}}$, at which the WDM transfer function is reduced by a factor of 2 relative to the corresponding CDM model with the same cosmological parameters. Explicitly, $M_{\mathrm{hm}}=\frac{4\pi}{3}\bar{\rho}\left(\frac{\pi}{k_{\mathrm{hm}}}\right)^3$.
For the equivalent thermal relic masses considered here, $m_{\mathrm{WDM}}=0.5$, $0.8$, $1.1$, and $1.5~\mathrm{keV}$, the corresponding half-mode masses are $13.7$, $2.86$, $0.99$, and $0.35$ $\times\ 10^{10}\,h^{-1}\ M_{\odot}$, respectively.

Having specified the ingredients entering $\sigma(M)$, we now turn to the other key quantity in Equation~(\ref{nv_f}), namely the halo formation time $z_\mathrm{f}$.
For all simulations across the analyzed redshift range ($z=0-$5), we calculate the halo formation time based on the main branch of the subhalo merger tree. Specifically, for a halo identified at a given redshift $z$, its formation time $z_\mathrm{f}$ is defined as the redshift at which its main progenitor first reached half of its virial mass $M_{\mathrm{vir}} (M_{\mathrm{200}c})$ at redshift $z$. We exclude halos whose progenitors were once satellites, following a method similar to \citet{2022MNRAS.509.3441A}. This primarily concerns low-mass halos that undergo flybys and exhibit anomalously high concentrations. Additionally, for a subset of low-mass halos whose scale radii are comparable to the softening length, fits to the NFW profile yield unsatisfactory results—specifically, the derived concentrations are anomalously large (L. Gao 2026, private communication). 
To ensure the reliability of our concentration estimates, we remove such problematic low-mass halos from our subsequent analysis.

After this selection, each halo in our final sample has both a reliable concentration measurement and an associated formation peak height. We then characterize the statistical distributions of these quantities in narrow mass bins.
As introduced in Section~\ref{sec:intro}, the probability distribution functions (pdfs) of both halo concentration and formation time exhibit lognormal characteristics. This statistical behavior provides a crucial clue for deriving a universal relationship between the PVs of concentration and formation peak height, particularly for halos within constrained mass intervals.
Therefore, we fit a lognormal function to the pdfs of $c_\mathrm{vir}$ or $\nu_\mathrm{f}$ for halos within a set of mass bins. These pdfs are characterized by the PVs, denoted as $c_\mathrm{peak}$ and $\nu_\mathrm{f,peak}$, respectively. 

Figure~\ref{para_define} illustrates an example of deriving the PVs of concentration and formation peak height from the VSMDPL simulation at $z=0$. 
As can be seen from the figure, the lognormal function fits well with both distributions. 
The distributions show mass-dependent widths: Concentration scatters more (formation peak height scatters less) for low-mass halos, reversing for high-mass systems.

Since our model is calibrated using the PV rather than the median or mean concentration, Table~\ref{compare_c} quantifies the differences among these statistical measures for halos at several characteristic masses and redshifts.
Over a broad range of halo masses and redshifts, these concentrations consistently follow the previously noted ordering: mode $<$ median $<$ mean. As shown in the last two columns, the differences gradually vanish at larger halo masses.

\begin{table*}
\centering
\small
\setlength{\tabcolsep}{1.5pt}
\renewcommand\arraystretch{1.0}
\caption{Concentrations and Relative Differences for Characteristic Halo Masses}
\begin{tabular*}{\textwidth}{@{\extracolsep{\fill}} l c c c c c c c c c c c c c c c } 
\hline\hline 
$M_{\mathrm{vir}}$ 
& \multicolumn{3}{c}{$c_{\mathrm{peak}}$} 
& \multicolumn{3}{c}{$c_{\mathrm{median}}$} 
& \multicolumn{3}{c}{$c_{\mathrm{mean}}$}  
& \multicolumn{3}{c}{$\frac{\Delta c}{c}(c_{\mathrm{peak}},c_{\mathrm{median}})$} 
& \multicolumn{3}{c}{$\frac{\Delta c}{c}(c_{\mathrm{peak}},c_{\mathrm{mean}})$} \\ 
\cline{2-4} \cline{5-7} \cline{8-10} \cline{11-13} \cline{14-16}
$(h^{-1}M_{\odot})$ 
& $z=0$ & $z=1$ & $z=2$ & $z=0$ & $z=1$ & $z=2$ & $z=0$ & $z=1$ & $z=2$ & $z=0$ & $z=1$ & $z=2$ & $z=0$ & $z=1$ & $z=2$ \\
\hline
$1.00 \times 10^9$  & 17.69 & 9.91 & 6.71 & 20.92 & 10.67 & 7.00 & 23.10 & 11.32 & 7.16 & 0.18 & 0.08 & 0.04 & 0.31 & 0.14 & 0.07 \\
$1.00 \times 10^{11}$ & 13.11 & 8.05 & 5.52 & 13.56 & 8.15 & 5.62 & 13.95 & 8.22 & 5.67 & 0.03 & 0.01 & 0.02 & 0.06 & 0.02 & 0.03 \\
$1.00 \times 10^{13}$ & 9.67 & 5.70 & 4.20 & 9.66 & 5.74 & 4.18 & 9.71 & 5.77 & 4.22 & $\sim 0.0$ & 0.01 & $\sim 0.0$ & $\sim 0.0$ & 0.01 & $\sim 0.0$ \\
$1.00 \times 10^{14}$ & 6.09 & 3.92 & 4.08 & 6.13 & 3.99 & 4.18 & 6.15 & 4.16 & 4.49 & 0.01 & 0.02 & 0.02 & 0.01 & 0.06 & 0.10 \\
\hline
\end{tabular*}
\label{compare_c}
\end{table*}

\section{results} \label{results}

\subsection{Universal Fitting Formulae for Peak Concentration} \label{universal_model}

\begin{figure}
\centering
\includegraphics[width=8.0cm,height=8.0cm]{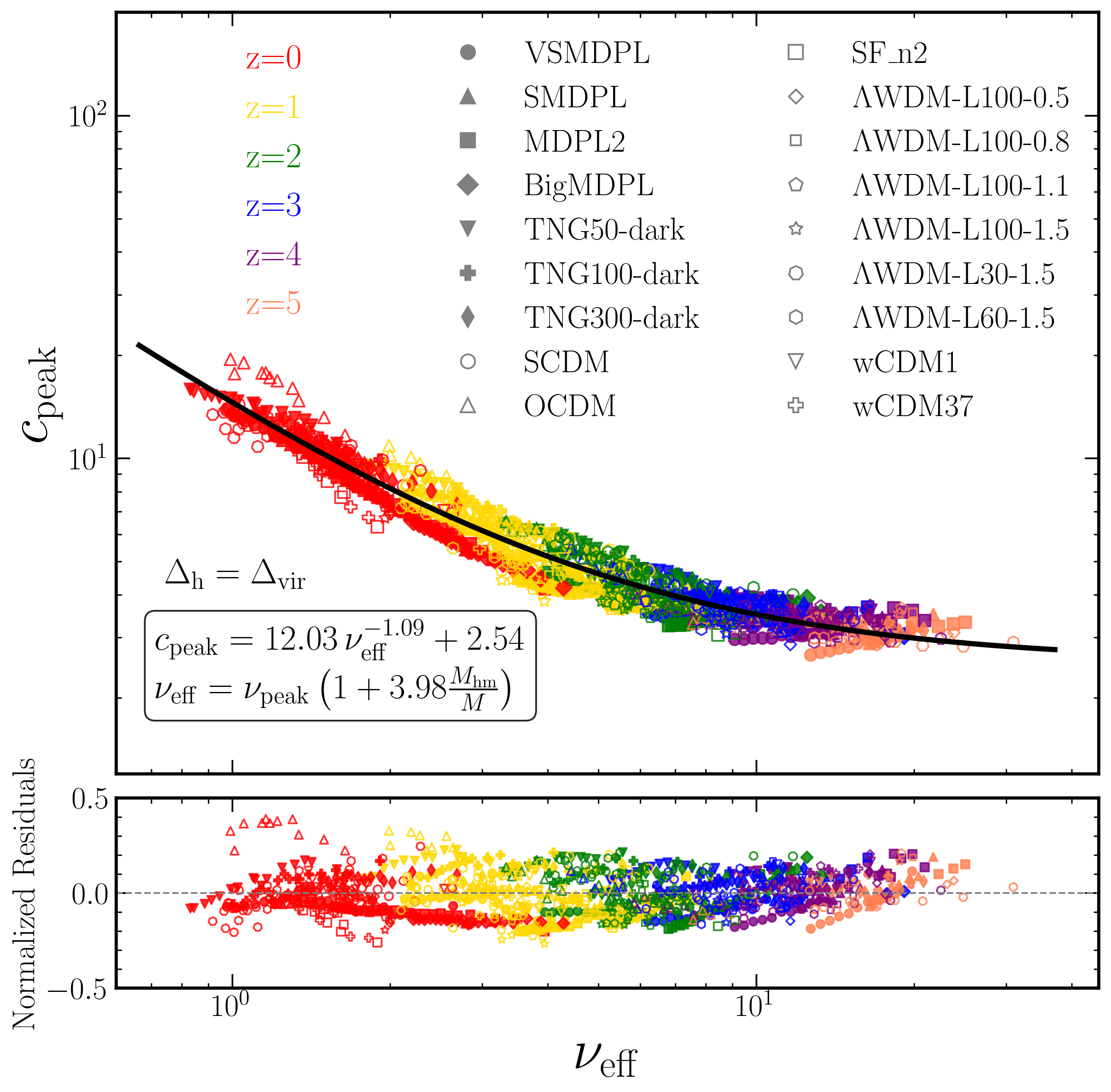}
\includegraphics[width=8.0cm,height=8.0cm]{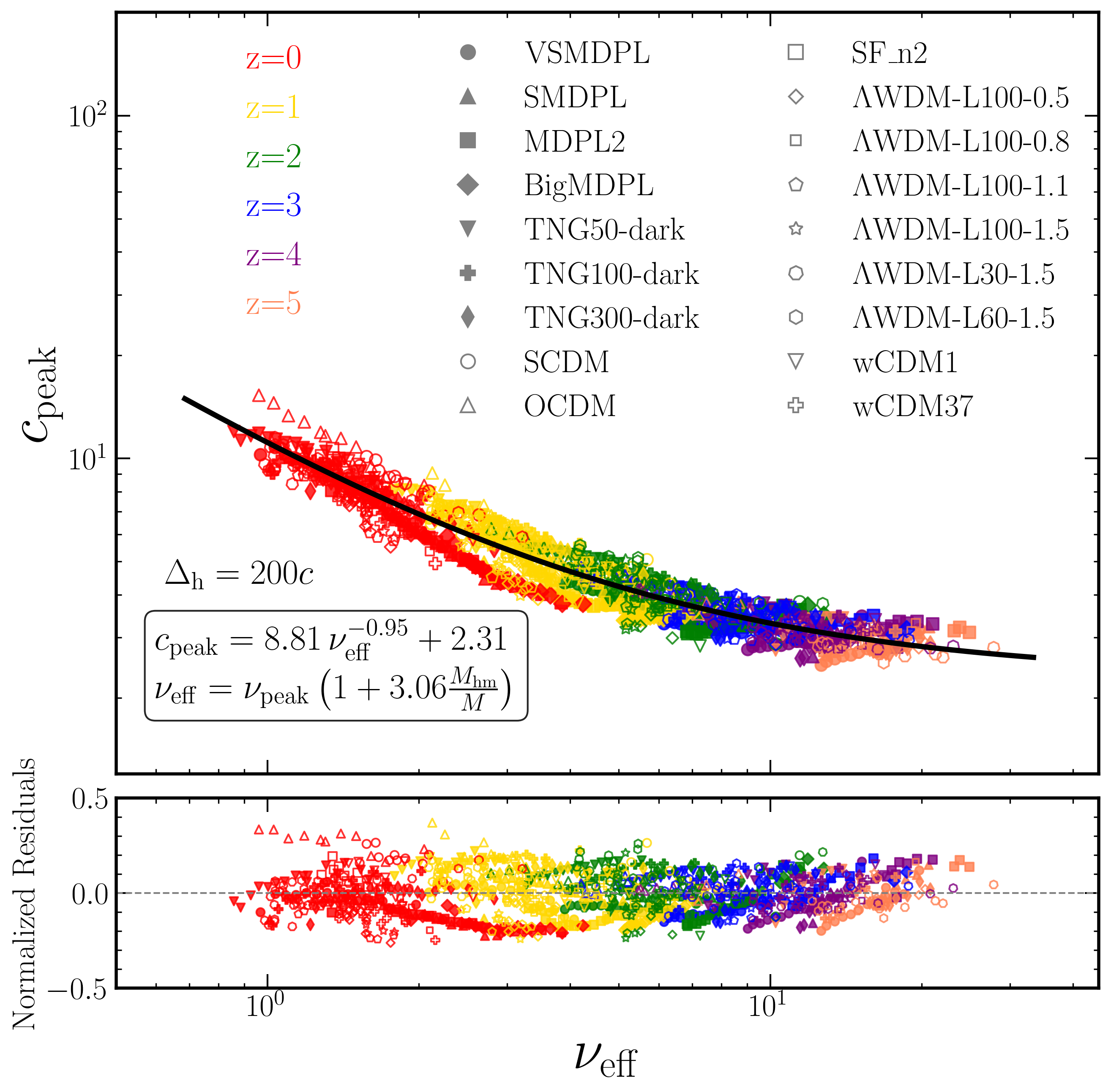}
\caption{
Universal fitting relations between $c_{\mathrm{peak}}$ and $\nu_{\mathrm{eff}}$ are shown for cosmological simulations across redshifts $z=$ 0--5. The left panel presents results for halos identified with $\Delta_{h} = \Delta_{\mathrm{vir}}$, while the right panel corresponds to $\Delta_{h} = 200c$. Point shapes denote different simulations, and colors indicate redshift, revealing significant data overlap. The solid lines show the best-fit relations given by Equations (\ref{c_nu_eq}) and (\ref{c_nu_eq_200}) for the left and right panels, respectively.}
\label{c_nu_peak}
\end{figure}

To characterize the relationship between halo structure and formation epoch, and to account for the suppression of low-mass halo formation in $\Lambda$WDM cosmologies, we define an effective formation peak height as
\begin{equation}
\nu_{\mathrm{eff}}=\nu_{\mathrm{f,peak}}\left(1+a\frac{M_{\mathrm{hm}}}{M}\right).
\end{equation}
For cosmologies without a WDM-like cutoff in the linear power spectrum, this correction is absent and $\nu_{\mathrm{eff}}$ reduces to $\nu_{\mathrm{f,peak}}$.
We present the PVs of halo concentration and formation peak height for all halos across the simulation suite in Figure~\ref{c_nu_peak}, where data for halos from different simulations are represented by distinct point markers and redshifts distinguished by color. 
The $c_\mathrm{{peak}}$--$\nu_\mathrm{{eff}}$ scaling relation derived from all simulations is well described by a power-law function of the form
\begin{equation}
c_\mathrm{{peak}}=12.03~ \nu_\mathrm{{eff}}^{-1.09}+2.54,
\label{c_nu_eq}
\end{equation}
with the best-fit value $a=3.98$. This fitting function, with only three free parameters (two coefficients and one power-law index) determined by least-squares fitting, accurately matches the results of all simulations (see the solid line in Figure~\ref{c_nu_peak}).

As shown in the bottom panel of Figure~\ref{c_nu_peak}, most normalized residuals remain below 20\%.  
The main exceptions are the OCDM data points (open triangles) at $\nu_\mathrm{eff} \lesssim 3$ (20\%$-$50\%).
The deviation in the OCDM results likely arises from the nonflat geometry, which causes the outer regions of halos to experience prominent cosmic expansion, leading to a higher density contrast between the inner and outer regions (i.e., elevated concentration). In addition, there are slightly positive deviations from the fitting line for results at large formation peak heights (i.e., for high-redshift halos and/or massive halos). Further discussion is provided later.

Here, we emphasize that the scatter is remarkably small, with negligible differences between simulations run under different cosmologies and with varying box sizes. Specifically, the dependence of concentration on halo mass, redshift, and cosmological parameters -- including the WDM suppression scale, characterized by $M_{\mathrm{hm}}-$is inherently captured by the simple $\nu_\mathrm{eff}$ ($\nu_\mathrm{f,peak}$) scaling described in Equation (\ref{c_nu_eq}).

As shown in the right panel of Figure~\ref{c_nu_peak}, we also explore the $c_{\mathrm{peak}}-\nu_{\mathrm{eff}}$ scaling relation using halos identified with an overdensity of $\Delta_{h} = 200c$. Similar to the results based on the virial overdensity ($\Delta_{h} = \Delta_{\mathrm{vir}}$) described by Equation (\ref{c_nu_eq}), the relation for $\Delta_{h} = 200c$ maintains a tight, universal behavior invariant to redshift, box size, and cosmology, exhibiting remarkably small scatter. Both relations are well characterized by a power-law function that flattens systematically with increasing effective formation peak height, approaching a constant concentration floor. However, quantitative differences emerge due to the adopted halo definition. Compared to Equation (\ref{c_nu_eq}), the best-fit relation derived for $\Delta = 200c$ is flatter and has a lower overall, amplitude:
\begin{equation}
c_{\mathrm{peak}} = 8.81 \nu_{\mathrm{eff}}^{-0.95} + 2.31,
\label{c_nu_eq_200}
\end{equation}
with the best-fit value $a=3.06$.
Furthermore, this relation yields a slightly lower concentration floor (2.31), demonstrating that while the universality of the scaling relation is robust, its exact shape and parameters depend on the chosen overdensity threshold. Given their identical functional form, we hereafter focus exclusively on Equation (\ref{c_nu_eq}).

To explicitly capture the dependencies and facilitate comparison with previous studies, we rewrite the formula by substituting the definition of $\nu_\mathrm{eff}$, yielding: 
\begin{equation}
\label{c_nu_eq2}
    c_\mathrm{{peak}}=6.81 [D(z_\mathrm{f,peak}) D(z)  \sigma(M) A_{\mathrm{hm}}]^{1.09}+2.54,
\end{equation}
where $A_{\mathrm{hm}}=(1+3.98 M_{\mathrm{hm}}/M)^{-1}$, and $z_\mathrm{f,peak}$ is determined for halos of mass $M$ at redshift $z$ (see Section \ref{method}), and thus depends on both $M$ and $z$.
Consequently, the formula contains only two independent parameters: redshift $z$ and halo mass $M$ in non-$\Lambda$WDM cosmologies, while for $\Lambda$WDM it additionally depends on the half-mode mass $M_{\mathrm{hm}}$. 

The above formula demonstrates that concentration anticorrelates with $z$ and $M$ and exhibits several interesting properties. First, mathematically, if $z$ or $M$ $\to \infty$, then $D(z)$ or $\sigma(M)$ $\to 0$, and the concentration parameter converges to 2.54.
This minimum value is a fitted parameter determined empirically from simulation data, rather than a fundamental constant predicted a priori \citep[e.g.,][]{2014MNRAS.441..378L}.
Second, when neglecting $D(z_\mathrm{f,peak})$
and the correction factor $A_{\mathrm{hm}}$ for $\Lambda$WDM cosmologies, the formula simplifies to the previously derived $c$--$\nu$ relation, which predicts that halos with higher peak heights exhibit lower concentrations while maintaining a floor value.
If we further fix the redshift $z$, the formula reduces to the traditional $c$--$M$ relation (see below for discussion). Notably, it is easy to understand that the scatter in our fitting formula should be smaller than that in the traditional $c$--$\nu$ or $c$--$M$ relations, because of the inclusion of $D(z_\mathrm{f,peak})$ and the additional $\Lambda$WDM suppression encoded by $M_{\mathrm{hm}}$. 

We observe a marginally increasing concentration trend for halos with high formation peak heights in Figure~\ref{c_nu_peak}, albeit far subtler than previously reported trends in the $c$--$M$ relation (e.g., \citealp{2011ApJ...740..102K, 2012MNRAS.423.3018P, 2015ApJ...799..108D}).
Comparisons within the MultiDark-Planck simulation suite indicate that the concentration measurements from different resolution runs are broadly consistent in their overlapping mass ranges, with only modest residual offsets. 
This is also consistent with the comparative results at varying resolutions shown in Figure 4 of \citet{2007MNRAS.381.1450N}, which demonstrate that numerical resolution can affect halo concentration measurements.
Given the subtlety of the upturn and the known sensitivity of halo concentrations to numerical resolution, we regard this feature as potentially affected by residual resolution effects, rather than as a robust physical signal.
Nevertheless, further investigation of such phenomena is valuable for uncovering clues about structure formation.

Important caveat: At first glance, the formula might suggest that earlier-formed halos have lower concentrations, due to its positive dependence on $D(z_\mathrm{f,peak})$. 
However, this is incorrect, because in this paper we aim to study the peak concentration within each mass bin, and thus $z_\mathrm{f,peak}$ is not a tunable free parameter but rather a function of $M$ and $z$.
For the analytical prediction of $D(z_\mathrm{f,peak})$ derived from the excursion set model, refer to the Appendix.

\subsection{Fitting Formulae for $D(z_\mathrm{f,peak})$}

\begin{figure*}
\centering
 \includegraphics[width=18.0cm,height=11.0cm]{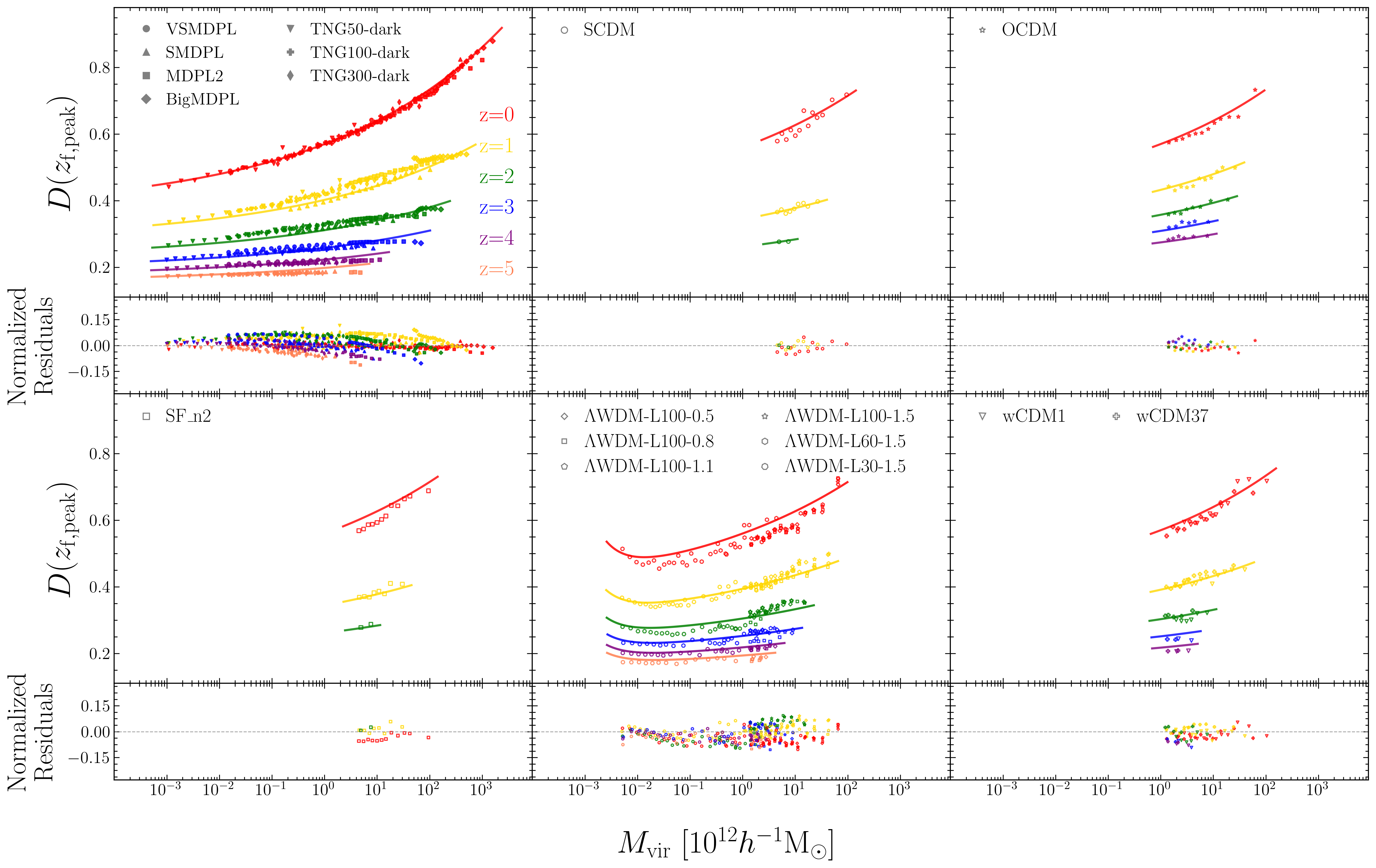}
\caption{The $D(z_{\mathrm{f,peak}})$--$M_{\mathrm{vir}}$--$\mathrm{redshift}$ relations are shown, where the solid lines represent Equation~(\ref{Dzf}) for halos identified with $\Delta_{h}=\Delta_{\mathrm{vir}}$.}
\label{FigDzf}
\end{figure*}

We previously showed that $D(z_\mathrm{f,peak})$ is a function of halo mass and redshift, with the corresponding analytical solution provided in the Appendix. However, owing to the substantial discrepancy between the model predictions and numerical simulation results for halo formation times, this analytical solution is not quantitatively accurate and serves only as a theoretical benchmark. 
Therefore, we derive a universal fitting formula, expressed as
\begin{equation}
\label{Dzf}
 D(z_\mathrm{f,peak}) \approx D(z) [0.20 M_{12}^{0.12} +0.36 D(z)^{-0.43}]\,\mathcal{E}(M,M_{\mathrm{hm}}),
\end{equation}
where $M_{12}=M/10^{12}\ h^{-1}\ M_{\odot}$.
The fits were actually performed using $\Lambda$CDM datasets, since these datasets span a wide mass range. Surprisingly, it turns out that the formula can also fit the results from other cosmological models. 
Similar to the introduction of $\nu_{\mathrm{eff}}$ in the previous subsection, we incorporate a correction factor $\mathcal{E}(M,M_{\mathrm{hm}})$ to capture the additional small-scale suppression in $\Lambda$WDM cosmologies. For halos identified with $\Delta_{h}=\Delta_{\mathrm{vir}}$, we adopt $\mathcal{E}(M,M_{\mathrm{hm}})=1+0.12(M_{\mathrm{hm}}/M)^{1.09}$. For cosmologies without a WDM-like cutoff in the linear power spectrum, we simply set $\mathcal{E}(M,M_{\mathrm{hm}})=1$.
Figure~\ref{FigDzf} shows the fitting results, with all relative errors within 15\%.
Similarly, for halos defined with an overdensity of $\Delta_{h} = 200c$, the corresponding fitting formula is
\begin{equation}
\label{Dzf_200c}
D(z_{\mathrm{f,peak}}) \approx D(z)[0.19 M_{12}^{0.12} + 0.36 D(z)^{-0.46}]\,\mathcal{E}(M,M_{\mathrm{hm}}),
\end{equation}
where we adopt $\mathcal{E}(M,M_{\mathrm{hm}})=1+0.13(M_{\mathrm{hm}}/M)^{1.05}$. This relation maintains the same functional form as Equation~(\ref{Dzf}), with only slightly different coefficients reflecting the adopted halo definition. Given this structural similarity, we omit the $\Delta_{h} = 200c$ results in Figure~\ref{FigDzf} and exclusively plot the $\Delta_{h} = \Delta_{\mathrm{vir}}$ case.
To derive a more accurate fitting formula, a more comprehensive suite of non-$\Lambda$ simulations spanning a broad mass range is required. However, such simulations are computationally intensive and will therefore be implemented in subsequent work.

\subsection{Interpretation and Comparison} \label{comparison_model}

With Equations~(\ref{c_nu_eq2}) and (\ref{Dzf}), we can predict the peak concentration, interpret the $c$--$M$ and $c$--$\nu$ relations, and compare with previous results.

\subsubsection{The $c$--$\nu$ Relation}
\label{c-nu}
The $c$--$\nu$ relation can be described by Equation (\ref{c_nu_eq}). 
Neglecting the factors $D(z_\mathrm{f,peak})^{-1}$ and $A_{\mathrm{hm}}^{-1}$, 
the equation simplifies to $c_\mathrm{peak}= b~\nu(M,z)^{-1.09}+2.54$, 
where $b$ is a coefficient that is, in fact, a function of $z$ and $M$.
However, if we apply the analytical formula given by Equation (\ref{A4}) to calculate $D(z_\mathrm{f,peak})$, the equation becomes
$c_\mathrm{peak}= 12.03 \left[D(z)^2 /(1+0.44D(z) \Delta(M))A_{\mathrm{hm}}\right]^{1.09} \nu(M,0)^{-1.09}+2.54$, indicating a complicated $c$--$\nu$ relation, unless the term $1+0.44D(z) \Delta(M)$ can be simplified to either 1 or $0.44D(z) \Delta(M)$.
Furthermore, if the fitting formula in Equation (\ref{Dzf}) is applied, the $c$--$\nu$ relation becomes even more complicated.
These results are at odds with previous findings that simply assumed $c$ to be a joint power-law function of $D(z)$ and $\nu$ (e.g., \citealp{2013ApJ...766...32B}; \citealp{2014MNRAS.441.3359D}). 
For instance, \citet{2013ApJ...766...32B} fit their the simulation results via the relation $c_\mathrm{{vir}} \approx 7.7 D(z)^{0.9} \nu(M,0)^{-0.29}$, which exhibits no concentration floor at large $\nu$ and an amplitude variation of $\sim$30\% for massive cluster halos. They also found cosmological model dependence and concluded the relation was nonuniversal. 
\cite{2014MNRAS.441..378L} also presented a fitting formula for the median $c$--$\nu$ relation for cold mark matter halos at redshifts $0\leq z\leq 2$, but did not include results for halos at higher redshifts.

\subsubsection{The $c$--$M$ Relation}
\label{c-M}
In the following, we use Equation~(\ref{c_nu_eq2}) to interpret the $c$--$M$ relation and its dependencies on redshift.

First, at fixed halo mass, the redshift dependence of the concentration is primarily encoded in the linear growth factor $D(z)$, with modulation by $D(z_\mathrm{f,peak})$. 
Specifically, neglecting $D(z_\mathrm{f,peak})$, the concentration follows a power law $a^{1.09}$ in EdS or high-redshift $\Lambda$CDM universes, where the linear growth factor scales as $D(a) \propto a=\ 1/(1+z)$.
In other cosmological models, $D(a)$ also exhibits a redshift-dependent behavior that decreases toward higher redshifts. In general, with the $D(z_\mathrm{f,peak})$ term, the dependence of concentration on $a$ remains similar but with variations in power-law index. This naturally explains the previously observed trend of decreasing halo concentrations at higher redshifts. 
Clearly, the assumption of $c\propto a$ in some previous analytical models (e.g., \citealp{2001MNRAS.321..559B}; \citealp{2002ApJ...568...52W}; \citealp{2007MNRAS.381.1450N}) is too simple. Since $D(z)$ and $D(z_\mathrm{f,peak})$ inherently depend on cosmological parameters, the cosmological dependence of the concentration is partly reflected in its redshift dependence. However, there is also a dependence on the power spectrum. Therefore, for clarity, we present the cosmological dependence separately in Section~\ref{Cos-dep}.

Second, at fixed redshift $z$, the dependence of $c$ on $M$ is rooted in the combination of $D(z_\mathrm{f,peak})$ and $\sigma(M)$, which gives $c \propto M^{-\alpha}$, with $\alpha$ being a positive, mass-dependent parameter. 
In general, for halos with increasing mass, both $z_\mathrm{f,peak}$ and $\sigma(M)$ decrease. 
While $D(z_\mathrm{f,peak})$ shows a modest increase (see Equation~(\ref{Dzf})), this variation is minor.
Therefore, the concentration depends primarily on $\sigma(M)$, yielding the well-known result that more massive halos have lower concentrations in simulations with CDM models.
Specifically, for scale-free dark matter cosmologies, where the variance of the linear density field scales as $\sigma(M) \propto M^{-(n_s+3)/6}$,
the contribution of $\sigma(M)$ to concentration parameter follows a power-law relation $c \propto M^{-1.09(n_s+3)/6} \sim M^{-0.18(n_s+3)}$. In particular, for the SF-n2 simulation ($n_s\ =\ -2$), $\alpha \sim 0.18$.
In simulations with $\Lambda$CDM models, for halos with masses of $\sim 10^{11}-10^{14}\ h^{-1}\ M_{\odot}$, the slope of the $\sigma(M)$ relation is broadly consistent with that of the SF-n2 case, while flattening in lower-mass and steepening in higher-mass ones.
However, the slope ($\alpha$) of $-0.18$ appears to be steeper than that found in previous studies using $\Lambda$CDM simulations (typically $\alpha \sim 0.10$ at $z=0$).
The steepening can be mitigated by the inclusion of contributions from $D(z_\mathrm{f,peak})$, which introduces a slight positive correlation with halo mass, as shown in Equation~(\ref{Dzf}).
This factor counterbalances the otherwise excessively steep slope, yielding a slope of $-$0.06 (calculated as $-$0.18 + 0.12), a value in general agreement with the $\Lambda$CDM simulation results for halos at $z=0$ \citep[e.g.,][]{2007MNRAS.378...55M,2008MNRAS.390L..64D}. 
However, in WDM cosmologies, the $\sigma(M)$ for small halos varies little with mass below a critical threshold, 
depending on the free-streaming scale set by the equivalent thermal relic mass. This positive $c$--$M$ correlation at low masses is 
a generic feature of WDM cosmologies, arising from the suppression of small-scale power by free streaming, which delays the formation of low-mass halos. 
This delay introduces a characteristic mass scale, close to the half-mode mass, at which the formation time and concentration reach a maximum; halos with both larger and smaller masses tend to form later and thus have lower concentrations, producing the turnover in the WDM $c$--$M$ relation.
In our model, this effect is accounted for through the half-mode mass $M_{\mathrm{hm}}$, both in the definition of $\nu_{\mathrm{eff}}$ and in the fitting formula for $D(z_\mathrm{f,peak})$.
Thus, halos spanning a broad mass range can share similar formation times and concentrations, which can help to explain previous WDM simulation results (e.g., \citealp{2013MNRAS.428..882M}; \citealp{2016MNRAS.460.1214L}).

Third, the redshift dependence of the $c$--$M$ relation's slope can be interpreted as the combination of all varying terms and the existence of floor. For halos at higher redshift, both $D(z)$ and $D(z_\mathrm{f,peak})$ become smaller, thereby reducing their contributions relative to the floor value.
Furthermore, high-redshift halos are relatively less massive, and the dependence of $\sigma(M)$ on $M$ flattens, particularly in simulations featuring a scale-dependent power spectrum characterized by reduced power on small scales. These factors lead to a flatter $c$--$M$ relation at higher redshift, as previously found in $\Lambda$CDM simulations (e.g., \citealp{2003ApJ...597L...9Z}; \citealp{2008MNRAS.387..536G}). This flattening is clearly observed in
panels (a)$-$(d) of Figures~\ref{FigCos} and \ref{Fig_com_model}.

\subsection{The Dependence on Cosmological Parameters}
\label{Cos-dep}

\begin{figure*}[htp]
\centering
 \includegraphics[width=18.0cm,height=7.5cm]{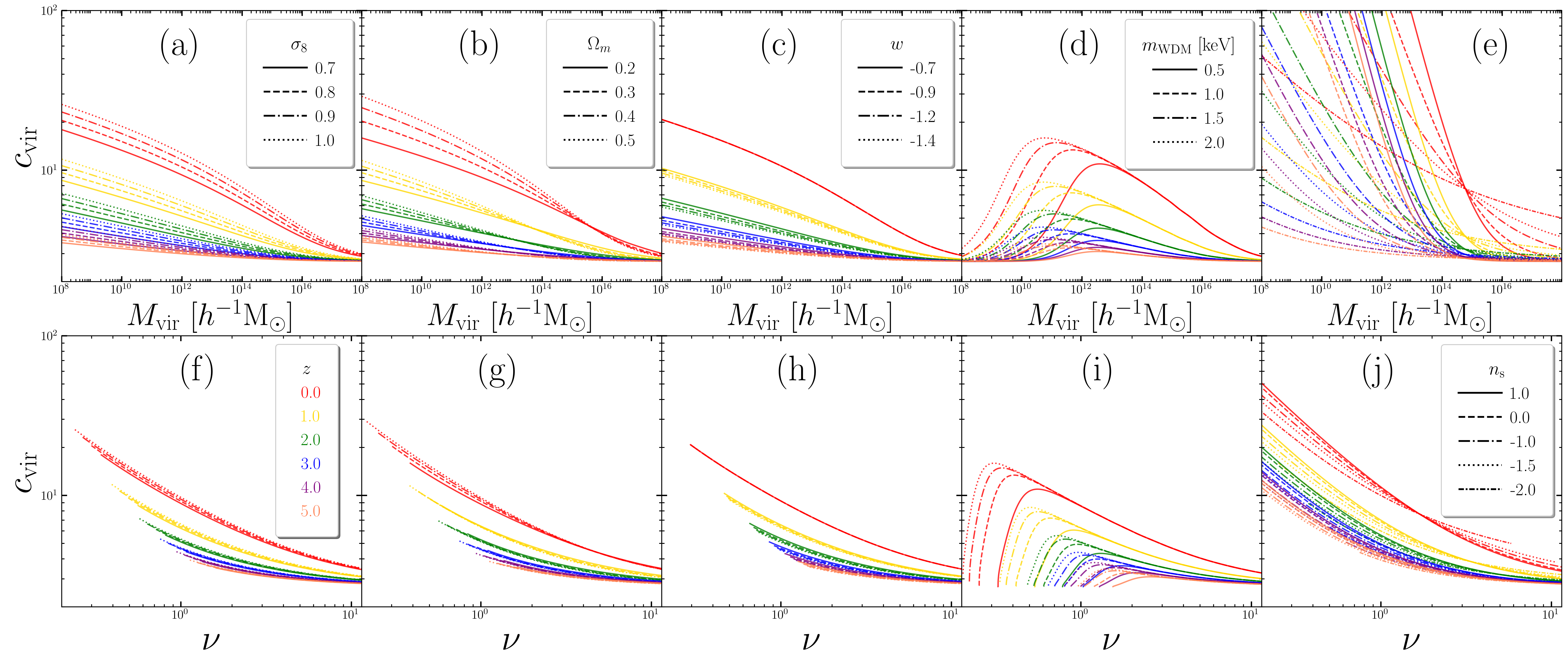}
\caption{The $c_{\mathrm{vir}}$-$M_{\mathrm{vir}}$ and $c_{\mathrm{vir}}$-$\nu$ relations are shown for various cosmological parameters ($\sigma_{8}$, $\Omega_{m}$, $w$, $m_{\mathrm{WDM}}$, and $n_s$), while the remaining parameters are fixed to the Planck18 values.}
\label{FigCos}
\end{figure*}

Within our framework, the cosmological dependence of halo concentration is primarily encapsulated in the linear growth factor and $\sigma(M)$, with an additional dependence on $M_\mathrm{hm}$ in the $\Lambda$WDM case.
Using Equations~(\ref{c_nu_eq2}) and (\ref{Dzf}), we predict and plot the peak concentrations of halos for different cosmological models in Figure~\ref{FigCos} to illustrate the $c$--$M$ and $c$--$\nu$ relations.

Panel (a) of the figure shows that halos of the same mass form with higher (lower) concentrations in simulations featuring elevated (suppressed) initial power-spectrum amplitude $\sigma_8$, since $\sigma(M) \propto \sigma_8$. 
panel (b) reveals that, in $\Lambda$CDM cosmological models with higher $\Omega_m$, halos have higher concentrations. These dependencies align with previous findings (e.g., \citealp{2008MNRAS.390L..64D}; \citealp{2014MNRAS.441.3359D}; \citealp{2016MNRAS.460.1214L}).
For the $w$CDM models (panel (c)), a smaller $|w|$ (which implies less dark energy) corresponds to a higher peak concentration, whereas results at $z=0$ show no such variation. This is because $w$ does not affect $\sigma(M)$ but influences the linear growth factor $D(z)$ (which is normalized to $z=0$), rendering its influence negligible at this redshift. Moreover, at other redshifts, the peak concentrations exhibit only mild variations  with changes in $w$.
It is interesting to see from panel (d), in 
$\Lambda$WDM models with smaller equivalent thermal relic masses (more suppression of small-scale density fluctuations),
that the halos in low-mass regimes have smaller concentrations and show a positive halo mass$-$concentration relation. This finding aligns with our previous interpretation and simulation results (e.g., \citealp{2013MNRAS.428..882M}; \citealp{2016MNRAS.460.1214L}). 
Our prescription even incorporates the shape of the power spectrum (via $\sigma(M)$). For instance, as indicated by Equation (\ref{c_nu_eq2}), simulations with enhanced small-scale power produce a steeper slope in the $c$--$M$ relation. The results for the scale-free spectrum are presented in panel (e) and are consistent with those presented in Figure 19 of \cite{2009ApJ...707..354Z}.\footnote{In \cite{2009ApJ...707..354Z}, the halo mass was normalized by the nonlinear mass scale $M_{\ast}$, defined implicitly by $\sigma(M_{\ast})=\delta_{sc}(z)$.} Note that a cross point exists at a halo mass of around $10^{15}\ h^{-1} \ M_{\odot}$, arising from the similarity in $\sigma(M)$ in this mass regime where the curves converge (see Figure 1 of \citealp{2009ApJ...707..354Z}).
Comparing the results in panels (a)–(d), we find that both the power-spectrum amplitude and matter density enhance halo concentrations. In contrast, dark energy delays structure formation through its impact on the linear growth factor, while warm dark matter also suppresses structure formation through its effect on small-scale fluctuations. These results are consistent with theoretical expectations.

Panels (f)-(j) of Figure~\ref{FigCos} present the results for the $c$--$\nu$ relation, which confirm the findings of \citet{2014MNRAS.441.3359D}, who reported a redshift-dependent form of this relation. In panel (i), the curves exhibit a turnover followed by a sharp decrease at low $\nu$.
We find that, while the definition of $\nu$ partially absorbs the dependence on cosmological parameters via $\delta_{sc}(z)$ (equivalently, $D(z)$) and $\sigma(M)$, it lacks information about the halo MAH (via $D(z_\mathrm{f,peak})$ in our framework), and thus cannot fully eliminate the redshift dependence to achieve a universal prescription.

\subsection{Comparison with Previous Models}
\label{Comp_Model}
\begin{figure*}
\centering
 \includegraphics[width=18.0cm,height=5.1cm]{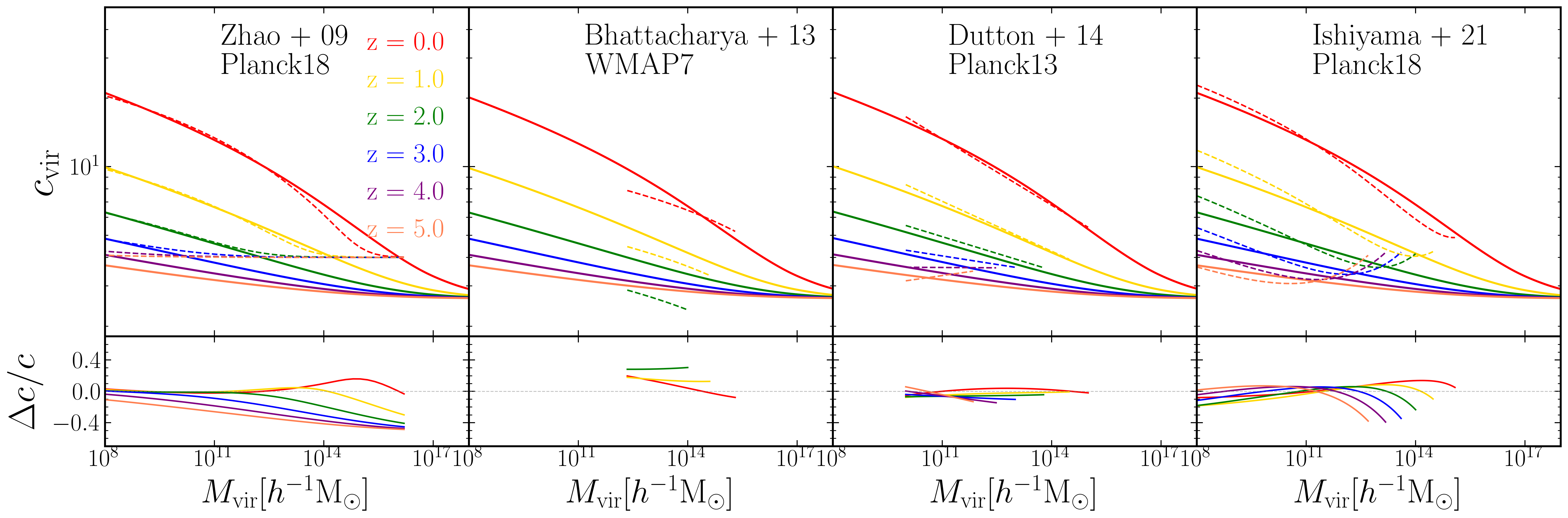}
 \includegraphics[width=9.5cm,height=5.1cm]{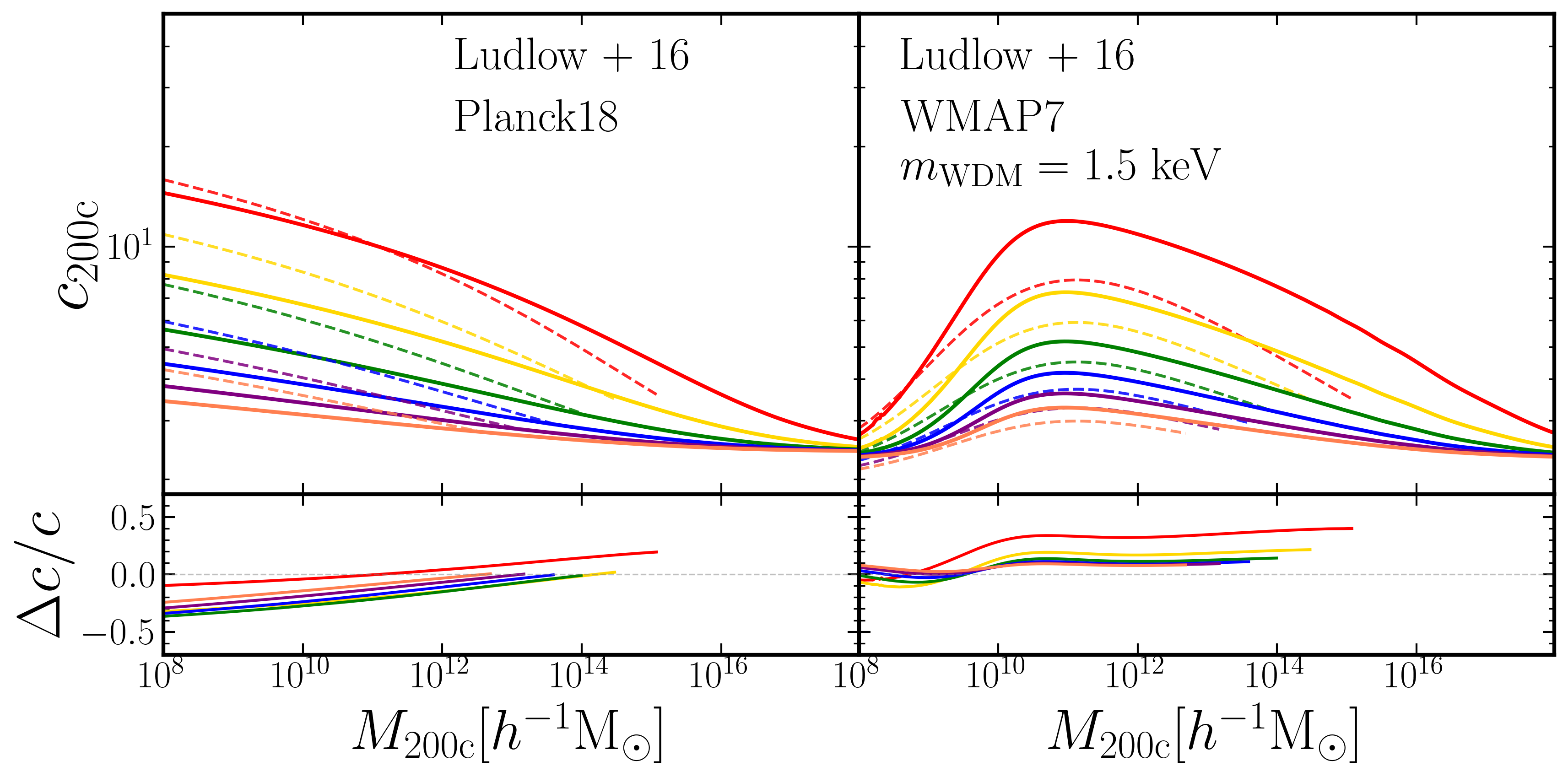}
\caption{
The $\Lambda$CDM and $\Lambda$WDM models in this work (solid lines) are compared with existing models from the literature (dashed lines) \citep{2009ApJ...707..354Z, 2013ApJ...766...32B, 2014MNRAS.441.3359D, 2016MNRAS.460.1214L, 2021MNRAS.506.4210I}.
The bottom subpanels show the relative differences between our results and those of the comparison models. 
The predictions are extended to very high halo masses to clearly illustrate the concentration floor of our model.}
\label{Fig_com_model}
\end{figure*}

The results shown in Figure~\ref{FigCos} indicate that the conventional $c$--$\nu$ relation (e.g., \citealp{2013ApJ...766...32B, 2014MNRAS.441.3359D}) is not fully universal across a wide range of redshifts and cosmological models. 
While $\nu(M,z)$ captures much of the dependence on halo mass, redshift, and cosmology, it does not encode halo assembly history and therefore cannot by itself eliminate the residual redshift dependence. 
In our framework, this missing information is carried by $D(z_{\mathrm{f, peak}})$, which motivates the introduction of the formation peak height. Similarly, for the $c$--$M$ relation, a power-law function also fails to provide a good fit to simulation results across a wide range of halo masses. 
The shape of this relation is indeed related to the shape of the primordial power spectrum, as previously analyzed. 
For example, the $c$--$M$ curve exhibits a peak in the results of WDM simulations.
Given our extensive early discussion, here we only compare our predictions with the $c$--$M$ relation from five representative models: \cite{2009ApJ...707..354Z}, \cite{2013ApJ...766...32B}, \cite{2014MNRAS.441.3359D}, \cite{2016MNRAS.460.1214L}, and \cite{2021MNRAS.506.4210I}. 
The results are shown in Figure~\ref{Fig_com_model}, with each subpanel plotting the relative differences between our predictions and those of a comparison model.
Please note that our predictions are for the peak concentration, which is smaller than the median value of a lognormal distribution.

In the model proposed by \citet{2009ApJ...707..354Z}, the median concentration is a function of the time elapsed since a halo accumulated 4\% of its mass, with a floor value of 4 (independent of halo mass and redshift). 
As shown in the first panel of Figure~\ref{Fig_com_model}, our results for peak concentration also exhibit a floor value of $\sim$2.54, located at the very high-mass end ($\sim$$10^{18}\ h^{-1}\ M_{\odot}$). 
Furthermore, the $c$--$M$ relation, calculated from our formula, flattens with increasing redshift, consistent with previous findings and our earlier argument.
The difference between our results and those of \citet{2009ApJ...707..354Z} becomes more pronounced toward the high-mass end.

The second panel presents the comparison to the predictions of \citet{2013ApJ...766...32B}, who considered a narrow mass range, and reveals significant differences. The third panel shows a comparison to the model of \citet{2014MNRAS.441.3359D}, where the differences appear small. However, \citet{2014MNRAS.441.3359D} predicted that the $c$--$M$ function flattens at high redshift but exhibits a positive slope at $z>4$, which is at odds with our results.

Following previous findings of an upturn in the $c$--$M$ relation (e.g., \citealp{2011ApJ...740..102K, 2012MNRAS.423.3018P, 2015ApJ...799..108D, 2016MNRAS.457.4340K}), the high-resolution Uchuu simulation also revealed a similar phenomenon, but only at $z>0.5$ \citep{2021MNRAS.506.4210I}. 
Such an upturn has sparked a persistent debate in the field. While \cite{2012MNRAS.427.1322L} attributed this phenomenon to dynamical state selection effects, \citet{2014MNRAS.441.3359D} argued that it could alternatively stem from the adoption of the $V_\mathrm{{max}}$ method to calculate halo concentrationS. Subsequently, \citet{2016MNRAS.457.4340K} proposed that this upturn represents an intrinsic feature of the $c$--$M$ relation, reasoning that elevated radial infall velocities in extremely massive halos enable accreted material to penetrate more deeply into the inner halo, consequently enhancing the concentration parameter. 
In the fourth panel, we compare our results with those of \citet{2021MNRAS.506.4210I} and find that the upturn deviates significantly from our results, with no agreement across the full range of redshifts and halo masses studied. 

Because the comparisons above are based primarily on $\Lambda$CDM models, we also compare our predictions with the concentration--mass model of \citet{2016MNRAS.460.1214L}, which was calibrated for both $\Lambda$CDM and $\Lambda$WDM cosmologies. Since their results are presented for halos defined by $\Delta_{h}=200c$, we use our corresponding fit for the same halo definition. The comparison is presented in the bottom two panels of Figure~\ref{Fig_com_model}. For $\Lambda$CDM, the model of \citet{2016MNRAS.460.1214L} generally predicts slightly higher concentrations than ours, except at $z=0$, where the agreement is closer. For $\Lambda$WDM with $m_{\mathrm{WDM}}=1.5\,\mathrm{keV}$, our model predicts systematically higher concentrations. 

The systematic offset between the two models likely arises from differences in methodology. \citet{2016MNRAS.460.1214L} relate concentration to the collapsed mass history and adopt an Einasto profile with fixed shape parameter $\alpha=0.18$, whereas our model is formulated in terms of the formation peak height and our concentrations are derived from NFW fits. In addition, for $\Lambda$WDM, the two models differ in the treatment of the fluctuation variance $\sigma(M)$, with our model adopting a sharp-$k$ filter. These differences are sufficient to account for the offset, especially in the $\Lambda$WDM case.

\section{Summary} \label{summary}

In this paper, we utilized a large suite of $N$-body simulations with diverse resolutions and cosmological models to derive fitting formulae for the most probable (peak) halo concentration parameter (of mass $M$ and redshift $z$). We revised the halo peak height parameter to the so-called formation peak-height, by incorporating the most probable formation epoch of these halos. Remarkably, we found that the formulae are universal and exhibit small scatter: The peak concentration can be determined solely and accurately by the formation peak height, with no explicit dependence on halo mass, redshift, cosmological model, or initial power spectrum. For $\Lambda$WDM halos, the additional effect of small-scale power suppression is captured by a correction associated with the half-mode mass $M_{\mathrm{hm}}$. The peak concentration exhibits a clear dependence on the joint product of the linear growth factor $D(z)$, the linear growth factor at the peak formation epoch $D(z_\mathrm{f,peak})$, and the rms of density fluctuation $\sigma(M)$. Notably, the fitting formulae are invariant to simulation conditions, including box size, resolution, and implementation code.

We used these universal fitting formulae to interpret the $c$--$M$ and $c$--$\nu$ relations reported in previous work, which exhibit intricate dependencies. Comparisons with our prescriptions yield consistent results and align well with the coherent physical picture inherent to the excursion set model. For instance, 
we recover an asymptotic concentration floor for high-mass halos, with the floor value determined empirically from the simulation data.
We further demonstrate the anticorrelations of concentration with redshift and mass, the positive correlations with halo mass (for small-mass halos in WDM simulations) and with the power-spectrum amplitude, and the dependence of the $c$--$M$ relation slope on the power-spectrum shape. 
Our predicted halo concentrations exhibit distinct dependencies on cosmological parameters, including $\sigma_{8}$, $\Omega_{m}$, and $m_{\mathrm{WDM}}$. If peak concentrations can be precisely measured via observational campaigns (e.g., lensing, galaxy or gas distributions, or kinematics, etc.), these empirical results could then be used to place constraints on the aforementioned cosmological parameters.

In conclusion, our fitting formulae are universal and can be readily used to predict the peak concentration of halos with mass $M$ at redshift $z$, given cosmological parameters and the power spectrum. Our definition of halo formation peak height incorporates exclusively the half-mass halo formation time, which is straightforward to determine. The linear growth factor at the peak halo formation epoch can also be calculated using our universal fitting formulae. For $\Lambda$WDM cosmologies, the prediction further depends on the half-mode mass $M_{\mathrm{hm}}$, which quantifies the scale of power-spectrum suppression. It can also be derived from semianalytical models or other fitting formulae for halo MAHs. A package for calculating the peak concentration is available on GitHub,\footnote{\url{https://github.com/wangdz630/concentration_universal_model/}.} with an archived version (D. Wang 2026) available on Zenodo (doi:\url{10.5281/zenodo.21090216}). The package also provides a detailed comparison of various cosmological models against the Planck cosmology predictions.

Finally, we emphasize that the dependencies in the $c$--$M$ and $c$--$\nu$ relations are so complex that deriving a universal prescription for these two correlations is nontrivial, and incorporating halo formation history emerges as a promising approach to this end. 
Notably, we focus solely on the average behavior of the concentration and do not investigate its dependence on halo formation time for same-mass halos$-$a focus of our ongoing work.

\begin{acknowledgments}

The authors thank the anonymous referee for constructive reports that greatly improved this paper.
We thank the TNG, MultiDark-Planck project and Dr. Behrooz for providing simulation data, and L. Gao for helpful discussion and comments. 
This research has been supported by National Natural Science Foundation of
China (grants Nos. 12573007, and 12073089).
Parts of the simulations and all of the analysis in this work were carried out at the Kunlun \& Loong HPC facility of the School of Physics and Astronomy, Sun Yat-sen University.
\end{acknowledgments}

\appendix
\section{Analytical solution of the linear growth factor at the peak formation epoch}
\label{appendix}
According to the EPS formalism \citep{1993MNRAS.262..627L}, the pdf of the scaled halo formation epoch,
\begin{equation}
\label{A1}
    \omega_\mathrm{f}=\frac{\delta_{sc}(z_\mathrm{f})-\delta_{sc}(z)}{\sqrt{\sigma^2(f M)-\sigma^2(M)}}
\end{equation}
can be written as
\begin{equation}
\label{A2}
    p(\omega_\mathrm{f})=2 \omega_\mathrm{f}~ \mathrm{erfc}(\omega_\mathrm{f}/\sqrt{2}). 
\end{equation}

The peak of the above pdf can be determined by solving $\frac{dp}{d\omega_\mathrm{f}}=0$, yielding $\omega_\mathrm{f,peak}\simeq 0.75$. 
Consequently, the critical overdensity at this peak is given by
\begin{equation}
\label{A3}
\delta_{sc}(z_\mathrm{f,peak}) \approx \delta_{sc}(z)+0.75~\Delta(M),
\end{equation}
where $\Delta(M)=\sqrt{\sigma^2(f M)-\sigma^2(M)}$.
Since $\delta_{sc}(z)=\delta_{sc}(0)/D(z)=1.686/D(z)$, substituting into Equation~(\ref{A3}) leads to
\begin{equation}
\label{A4}
    D(z_\mathrm{f,peak}) \approx \frac{D(z)}{1+\frac{0.75}{1.686}D(z)\Delta(M)}.
\end{equation}
Interestingly, when the second term in the denominator is $\ll 1$ (i.e., for massive halos), $D(z_\mathrm{f,peak})\approx D(z)$, independent of mass $M$. 
In the opposite regime, $D(z_\mathrm{f,peak})\approx2.25/\Delta(M)$ (i.e., for small halos), independent of redshift $z$.
As halos form slightly earlier in simulations than predicted by theory \citep[e.g.,][]{2003MNRAS.344.1327L}, the above formula with $f\approx0.16$ provides a better fit to the simulation results for $D(z_\mathrm{f,peak})$ than those with $f=1/2$. Therefore, for $f\approx0.16$, Equation~(\ref{A4}) may serve as a useful approximation to the simulation-inferred $D(z_\mathrm{f,peak})$, as given by Equations~(\ref{Dzf}) and~(\ref{Dzf_200c}), for predicting halo concentrations in both CDM and WDM cosmologies.

\bibliography{references}{}

@ARTICLE{1996ApJ...462..563N,
       author = {{Navarro}, Julio F. and {Frenk}, Carlos S. and {White}, Simon D.~M.},
        title = "{The Structure of Cold Dark Matter Halos}",
      journal = {\apj},
         year = 1996,
        month = may,
       volume = {462},
        pages = {563},
          doi = {10.1086/177173},
archivePrefix = {arXiv},
       eprint = {astro-ph/9508025},
 primaryClass = {astro-ph},
       adsurl = {https://ui.adsabs.harvard.edu/abs/1996ApJ...462..563N}
}

@ARTICLE{1997ApJ...490..493N,
       author = {{Navarro}, Julio F. and {Frenk}, Carlos S. and {White}, Simon D.~M.},
        title = "{A Universal Density Profile from Hierarchical Clustering}",
      journal = {\apj},
         year = 1997,
        month = dec,
       volume = {490},
       number = {2},
        pages = {493-508},
          doi = {10.1086/304888},
archivePrefix = {arXiv},
       eprint = {astro-ph/9611107},
 primaryClass = {astro-ph},
       adsurl = {https://ui.adsabs.harvard.edu/abs/1997ApJ...490..493N}
}

@ARTICLE{2000ApJ...535...30J,
       author = {{Jing}, Y.~P.},
        title = "{The Density Profile of Equilibrium and Nonequilibrium Dark Matter Halos}",
      journal = {\apj},
         year = 2000,
        month = may,
       volume = {535},
       number = {1},
        pages = {30-36},
          doi = {10.1086/308809},
archivePrefix = {arXiv},
       eprint = {astro-ph/9901340},
 primaryClass = {astro-ph},
       adsurl = {https://ui.adsabs.harvard.edu/abs/2000ApJ...535...30J}
}

@ARTICLE{2014MNRAS.441..378L,
       author = {{Ludlow}, Aaron D. and {Navarro}, Julio F. and {Angulo}, Ra{\'u}l E. and {Boylan-Kolchin}, Michael and {Springel}, Volker and {Frenk}, Carlos and {White}, Simon D.~M.},
        title = "{The mass-concentration-redshift relation of cold dark matter haloes}",
      journal = {\mnras},
         year = 2014,
        month = jun,
       volume = {441},
       number = {1},
        pages = {378-388},
          doi = {10.1093/mnras/stu483},
archivePrefix = {arXiv},
       eprint = {1312.0945},
 primaryClass = {astro-ph.CO},
       adsurl = {https://ui.adsabs.harvard.edu/abs/2014MNRAS.441..378L}
}

@ARTICLE{2016MNRAS.460.1214L,
       author = {{Ludlow}, Aaron D. and {Bose}, Sownak and {Angulo}, Ra{\'u}l E. and {Wang}, Lan and {Hellwing}, Wojciech A. and {Navarro}, Julio F. and {Cole}, Shaun and {Frenk}, Carlos S.},
        title = "{The mass-concentration-redshift relation of cold and warm dark matter haloes}",
      journal = {\mnras},
         year = 2016,
        month = aug,
       volume = {460},
       number = {2},
        pages = {1214-1232},
          doi = {10.1093/mnras/stw1046},
archivePrefix = {arXiv},
       eprint = {1601.02624},
 primaryClass = {astro-ph.CO},
       adsurl = {https://ui.adsabs.harvard.edu/abs/2016MNRAS.460.1214L}
}

@ARTICLE{2024MNRAS.52710760W,
       author = {{Wang}, Kai and {Mo}, H.~J. and {Chen}, Yangyao and {Schaye}, Joop},
        title = "{An efficient and robust method to estimate halo concentration based on the method of moments}",
      journal = {\mnras},
         year = 2024,
        month = feb,
       volume = {527},
       number = {4},
        pages = {10760-10776},
          doi = {10.1093/mnras/stad3927},
archivePrefix = {arXiv},
       eprint = {2310.00200},
 primaryClass = {astro-ph.CO},
       adsurl = {https://ui.adsabs.harvard.edu/abs/2024MNRAS.52710760W}
}

@ARTICLE{1998ApJ...495...80B,
       author = {{Bryan}, Greg L. and {Norman}, Michael L.},
        title = "{Statistical Properties of X-Ray Clusters: Analytic and Numerical Comparisons}",
      journal = {\apj},
         year = 1998,
        month = mar,
       volume = {495},
       number = {1},
        pages = {80-99},
          doi = {10.1086/305262},
archivePrefix = {arXiv},
       eprint = {astro-ph/9710107},
 primaryClass = {astro-ph},
       adsurl = {https://ui.adsabs.harvard.edu/abs/1998ApJ...495...80B}
}

@ARTICLE{2001MNRAS.321..559B,
       author = {{Bullock}, J.~S. and {Kolatt}, T.~S. and {Sigad}, Y. and {Somerville}, R.~S. and {Kravtsov}, A.~V. and {Klypin}, A.~A. and {Primack}, J.~R. and {Dekel}, A.},
        title = "{Profiles of dark haloes: evolution, scatter and environment}",
      journal = {\mnras},
         year = 2001,
        month = mar,
       volume = {321},
       number = {3},
        pages = {559-575},
          doi = {10.1046/j.1365-8711.2001.04068.x},
archivePrefix = {arXiv},
       eprint = {astro-ph/9908159},
 primaryClass = {astro-ph},
       adsurl = {https://ui.adsabs.harvard.edu/abs/2001MNRAS.321..559B}
}

@ARTICLE{2002ApJ...568...52W,
       author = {{Wechsler}, Risa H. and {Bullock}, James S. and {Primack}, Joel R. and {Kravtsov}, Andrey V. and {Dekel}, Avishai},
        title = "{Concentrations of Dark Halos from Their Assembly Histories}",
      journal = {\apj},
         year = 2002,
        month = mar,
       volume = {568},
       number = {1},
        pages = {52-70},
          doi = {10.1086/338765},
archivePrefix = {arXiv},
       eprint = {astro-ph/0108151},
 primaryClass = {astro-ph},
       adsurl = {https://ui.adsabs.harvard.edu/abs/2002ApJ...568...52W}
}

@ARTICLE{2014MNRAS.441.3359D,
       author = {{Dutton}, Aaron A. and {Macci{\`o}}, Andrea V.},
        title = "{Cold dark matter haloes in the Planck era: evolution of structural parameters for Einasto and NFW profiles}",
      journal = {\mnras},
         year = 2014,
        month = jul,
       volume = {441},
       number = {4},
        pages = {3359-3374},
          doi = {10.1093/mnras/stu742},
archivePrefix = {arXiv},
       eprint = {1402.7073},
 primaryClass = {astro-ph.CO},
       adsurl = {https://ui.adsabs.harvard.edu/abs/2014MNRAS.441.3359D}
}

@ARTICLE{2008MNRAS.390L..64D,
       author = {{Duffy}, Alan R. and {Schaye}, Joop and {Kay}, Scott T. and {Dalla Vecchia}, Claudio},
        title = "{Dark matter halo concentrations in the Wilkinson Microwave Anisotropy Probe year 5 cosmology}",
      journal = {\mnras},
         year = 2008,
        month = oct,
       volume = {390},
       number = {1},
        pages = {L64-L68},
          doi = {10.1111/j.1745-3933.2008.00537.x},
archivePrefix = {arXiv},
       eprint = {0804.2486},
 primaryClass = {astro-ph},
       adsurl = {https://ui.adsabs.harvard.edu/abs/2008MNRAS.390L..64D}
}

@ARTICLE{2013MNRAS.428..882M,
       author = {{Macci{\`o}}, Andrea V. and {Ruchayskiy}, Oleg and {Boyarsky}, Alexey and {Mu{\~n}oz-Cuartas}, Juan C.},
        title = "{The inner structure of haloes in cold+warm dark matter models}",
      journal = {\mnras},
         year = 2013,
        month = jan,
       volume = {428},
       number = {1},
        pages = {882-890},
          doi = {10.1093/mnras/sts078},
archivePrefix = {arXiv},
       eprint = {1202.2858},
 primaryClass = {astro-ph.CO},
       adsurl = {https://ui.adsabs.harvard.edu/abs/2013MNRAS.428..882M}
}

@ARTICLE{2001ApJ...554..114E,
       author = {{Eke}, Vincent R. and {Navarro}, Julio F. and {Steinmetz}, Matthias},
        title = "{The Power Spectrum Dependence of Dark Matter Halo Concentrations}",
      journal = {\apj},
         year = 2001,
        month = jun,
       volume = {554},
       number = {1},
        pages = {114-125},
          doi = {10.1086/321345},
archivePrefix = {arXiv},
       eprint = {astro-ph/0012337},
 primaryClass = {astro-ph},
       adsurl = {https://ui.adsabs.harvard.edu/abs/2001ApJ...554..114E}
}

@ARTICLE{2005MNRAS.357...82R,
       author = {{Reed}, Darren and {Governato}, Fabio and {Verde}, Licia and {Gardner}, Jeffrey and {Quinn}, Thomas and {Stadel}, Joachim and {Merritt}, David and {Lake}, George},
        title = "{Evolution of the density profiles of dark matter haloes}",
      journal = {\mnras},
         year = 2005,
        month = feb,
       volume = {357},
       number = {1},
        pages = {82-96},
          doi = {10.1111/j.1365-2966.2005.08612.x},
archivePrefix = {arXiv},
       eprint = {astro-ph/0312544},
 primaryClass = {astro-ph},
       adsurl = {https://ui.adsabs.harvard.edu/abs/2005MNRAS.357...82R}
}

@ARTICLE{2007MNRAS.378...55M,
       author = {{Macci{\`o}}, Andrea V. and {Dutton}, Aaron A. and {van den Bosch}, Frank C. and {Moore}, Ben and {Potter}, Doug and {Stadel}, Joachim},
        title = "{Concentration, spin and shape of dark matter haloes: scatter and the dependence on mass and environment}",
      journal = {\mnras},
         year = 2007,
        month = jun,
       volume = {378},
       number = {1},
        pages = {55-71},
          doi = {10.1111/j.1365-2966.2007.11720.x},
archivePrefix = {arXiv},
       eprint = {astro-ph/0608157},
 primaryClass = {astro-ph},
       adsurl = {https://ui.adsabs.harvard.edu/abs/2007MNRAS.378...55M}
}

@ARTICLE{2007MNRAS.381.1450N,
       author = {{Neto}, Angelo F. and {Gao}, Liang and {Bett}, Philip and {Cole}, Shaun and {Navarro}, Julio F. and {Frenk}, Carlos S. and {White}, Simon D.~M. and {Springel}, Volker and {Jenkins}, Adrian},
        title = "{The statistics of {\ensuremath{\Lambda}} CDM halo concentrations}",
      journal = {\mnras},
         year = 2007,
        month = nov,
       volume = {381},
       number = {4},
        pages = {1450-1462},
          doi = {10.1111/j.1365-2966.2007.12381.x},
archivePrefix = {arXiv},
       eprint = {0706.2919},
 primaryClass = {astro-ph},
       adsurl = {https://ui.adsabs.harvard.edu/abs/2007MNRAS.381.1450N}
}

@ARTICLE{2003MNRAS.339...12Z,
       author = {{Zhao}, D.~H. and {Mo}, H.~J. and {Jing}, Y.~P. and {B{\"o}rner}, G.},
        title = "{The growth and structure of dark matter haloes}",
      journal = {\mnras},
         year = 2003,
        month = feb,
       volume = {339},
       number = {1},
        pages = {12-24},
          doi = {10.1046/j.1365-8711.2003.06135.x},
archivePrefix = {arXiv},
       eprint = {astro-ph/0204108},
 primaryClass = {astro-ph},
       adsurl = {https://ui.adsabs.harvard.edu/abs/2003MNRAS.339...12Z}
}

@ARTICLE{1986ApJ...304...15B,
       author = {{Bardeen}, J.~M. and {Bond}, J.~R. and {Kaiser}, N. and {Szalay}, A.~S.},
        title = "{The Statistics of Peaks of Gaussian Random Fields}",
      journal = {\apj},
         year = 1986,
        month = may,
       volume = {304},
        pages = {15},
          doi = {10.1086/164143},
       adsurl = {https://ui.adsabs.harvard.edu/abs/1986ApJ...304...15B}
}

@ARTICLE{2009ApJ...707..354Z,
       author = {{Zhao}, D.~H. and {Jing}, Y.~P. and {Mo}, H.~J. and {B{\"o}rner}, G.},
        title = "{Accurate Universal Models for the Mass Accretion Histories and Concentrations of Dark Matter Halos}",
      journal = {\apj},
         year = 2009,
        month = dec,
       volume = {707},
       number = {1},
        pages = {354-369},
          doi = {10.1088/0004-637X/707/1/354},
archivePrefix = {arXiv},
       eprint = {0811.0828},
 primaryClass = {astro-ph},
       adsurl = {https://ui.adsabs.harvard.edu/abs/2009ApJ...707..354Z}
}

@ARTICLE{2015ApJ...799..108D,
       author = {{Diemer}, Benedikt and {Kravtsov}, Andrey V.},
        title = "{A Universal Model for Halo Concentrations}",
      journal = {\apj},
         year = 2015,
        month = jan,
       volume = {799},
       number = {1},
          eid = {108},
        pages = {108},
          doi = {10.1088/0004-637X/799/1/108},
archivePrefix = {arXiv},
       eprint = {1407.4730},
 primaryClass = {astro-ph.CO},
       adsurl = {https://ui.adsabs.harvard.edu/abs/2015ApJ...799..108D}
}

@ARTICLE{2019ApJ...871..168D,
       author = {{Diemer}, Benedikt and {Joyce}, Michael},
        title = "{An Accurate Physical Model for Halo Concentrations}",
      journal = {\apj},
         year = 2019,
        month = feb,
       volume = {871},
       number = {2},
          eid = {168},
        pages = {168},
          doi = {10.3847/1538-4357/aafad6},
archivePrefix = {arXiv},
       eprint = {1809.07326},
 primaryClass = {astro-ph.CO},
       adsurl = {https://ui.adsabs.harvard.edu/abs/2019ApJ...871..168D}
}

@ARTICLE{2003ApJ...597L...9Z,
       author = {{Zhao}, D.~H. and {Jing}, Y.~P. and {Mo}, H.~J. and {B{\"o}rner}, G.},
        title = "{Mass and Redshift Dependence of Dark Halo Structure}",
      journal = {\apjl},
         year = 2003,
        month = nov,
       volume = {597},
       number = {1},
        pages = {L9-L12},
          doi = {10.1086/379734},
archivePrefix = {arXiv},
       eprint = {astro-ph/0309375},
 primaryClass = {astro-ph},
       adsurl = {https://ui.adsabs.harvard.edu/abs/2003ApJ...597L...9Z}
}

@ARTICLE{2008MNRAS.387..536G,
       author = {{Gao}, Liang and {Navarro}, Julio F. and {Cole}, Shaun and {Frenk}, Carlos S. and {White}, Simon D.~M. and {Springel}, Volker and {Jenkins}, Adrian and {Neto}, Angelo F.},
        title = "{The redshift dependence of the structure of massive {\ensuremath{\Lambda}} cold dark matter haloes}",
      journal = {\mnras},
         year = 2008,
        month = jun,
       volume = {387},
       number = {2},
        pages = {536-544},
          doi = {10.1111/j.1365-2966.2008.13277.x},
archivePrefix = {arXiv},
       eprint = {0711.0746},
 primaryClass = {astro-ph},
       adsurl = {https://ui.adsabs.harvard.edu/abs/2008MNRAS.387..536G}
}

@ARTICLE{2008MNRAS.391.1940M,
       author = {{Macci{\`o}}, Andrea V. and {Dutton}, Aaron A. and {van den Bosch}, Frank C.},
        title = "{Concentration, spin and shape of dark matter haloes as a function of the cosmological model: WMAP1, WMAP3 and WMAP5 results}",
      journal = {\mnras},
         year = 2008,
        month = dec,
       volume = {391},
       number = {4},
        pages = {1940-1954},
          doi = {10.1111/j.1365-2966.2008.14029.x},
archivePrefix = {arXiv},
       eprint = {0805.1926},
 primaryClass = {astro-ph},
       adsurl = {https://ui.adsabs.harvard.edu/abs/2008MNRAS.391.1940M}
}

@ARTICLE{2011ApJ...740..102K,
       author = {{Klypin}, Anatoly A. and {Trujillo-Gomez}, Sebastian and {Primack}, Joel},
        title = "{Dark Matter Halos in the Standard Cosmological Model: Results from the Bolshoi Simulation}",
      journal = {\apj},
         year = 2011,
        month = oct,
       volume = {740},
       number = {2},
          eid = {102},
        pages = {102},
          doi = {10.1088/0004-637X/740/2/102},
archivePrefix = {arXiv},
       eprint = {1002.3660},
 primaryClass = {astro-ph.CO},
       adsurl = {https://ui.adsabs.harvard.edu/abs/2011ApJ...740..102K}
}

@ARTICLE{2013ApJ...766...32B,
       author = {{Bhattacharya}, Suman and {Habib}, Salman and {Heitmann}, Katrin and {Vikhlinin}, Alexey},
        title = "{Dark Matter Halo Profiles of Massive Clusters: Theory versus Observations}",
      journal = {\apj},
         year = 2013,
        month = mar,
       volume = {766},
       number = {1},
          eid = {32},
        pages = {32},
          doi = {10.1088/0004-637X/766/1/32},
archivePrefix = {arXiv},
       eprint = {1112.5479},
 primaryClass = {astro-ph.CO},
       adsurl = {https://ui.adsabs.harvard.edu/abs/2013ApJ...766...32B}
}

@ARTICLE{2012MNRAS.427.1322L,
       author = {{Ludlow}, Aaron D. and {Navarro}, Julio F. and {Li}, Ming and {Angulo}, Raul E. and {Boylan-Kolchin}, Michael and {Bett}, Philip E.},
        title = "{The dynamical state and mass-concentration relation of galaxy clusters}",
      journal = {\mnras},
         year = 2012,
        month = dec,
       volume = {427},
       number = {2},
        pages = {1322-1328},
          doi = {10.1111/j.1365-2966.2012.21892.x},
archivePrefix = {arXiv},
       eprint = {1206.1049},
 primaryClass = {astro-ph.CO},
       adsurl = {https://ui.adsabs.harvard.edu/abs/2012MNRAS.427.1322L}
}

@ARTICLE{2016MNRAS.457.4340K,
       author = {{Klypin}, Anatoly and {Yepes}, Gustavo and {Gottl{\"o}ber}, Stefan and {Prada}, Francisco and {He{\ss}}, Steffen},
        title = "{MultiDark simulations: the story of dark matter halo concentrations and density profiles}",
      journal = {\mnras},
         year = 2016,
        month = apr,
       volume = {457},
       number = {4},
        pages = {4340-4359},
          doi = {10.1093/mnras/stw248},
archivePrefix = {arXiv},
       eprint = {1411.4001},
 primaryClass = {astro-ph.CO},
       adsurl = {https://ui.adsabs.harvard.edu/abs/2016MNRAS.457.4340K}
}

@ARTICLE{1991ApJ...379..440B,
       author = {{Bond}, J.~R. and {Cole}, S. and {Efstathiou}, G. and {Kaiser}, N.},
        title = "{Excursion Set Mass Functions for Hierarchical Gaussian Fluctuations}",
      journal = {\apj},
         year = 1991,
        month = oct,
       volume = {379},
        pages = {440},
          doi = {10.1086/170520},
       adsurl = {https://ui.adsabs.harvard.edu/abs/1991ApJ...379..440B}
}

@ARTICLE{1993MNRAS.262..627L,
       author = {{Lacey}, Cedric and {Cole}, Shaun},
        title = "{Merger rates in hierarchical models of galaxy formation}",
      journal = {\mnras},
         year = 1993,
        month = jun,
       volume = {262},
       number = {3},
        pages = {627-649},
          doi = {10.1093/mnras/262.3.627},
       adsurl = {https://ui.adsabs.harvard.edu/abs/1993MNRAS.262..627L}
}

@ARTICLE{2000MNRAS.319..168C,
       author = {{Cole}, Shaun and {Lacey}, Cedric G. and {Baugh}, Carlton M. and {Frenk}, Carlos S.},
        title = "{Hierarchical galaxy formation}",
      journal = {\mnras},
         year = 2000,
        month = nov,
       volume = {319},
       number = {1},
        pages = {168-204},
          doi = {10.1046/j.1365-8711.2000.03879.x},
archivePrefix = {arXiv},
       eprint = {astro-ph/0007281},
 primaryClass = {astro-ph},
       adsurl = {https://ui.adsabs.harvard.edu/abs/2000MNRAS.319..168C}
}

@ARTICLE{2008MNRAS.383..546C,
       author = {{Cole}, Shaun and {Helly}, John and {Frenk}, Carlos S. and {Parkinson}, Hannah},
        title = "{The statistical properties of {\ensuremath{\Lambda}} cold dark matter halo formation}",
      journal = {\mnras},
         year = 2008,
        month = jan,
       volume = {383},
       number = {2},
        pages = {546-556},
          doi = {10.1111/j.1365-2966.2007.12516.x},
archivePrefix = {arXiv},
       eprint = {0708.1376},
 primaryClass = {astro-ph},
       adsurl = {https://ui.adsabs.harvard.edu/abs/2008MNRAS.383..546C}
}

@ARTICLE{2003MNRAS.344.1327L,
       author = {{Lin}, W.~P. and {Jing}, Y.~P. and {Lin}, Lihwai},
        title = "{Formation time-distribution of dark matter haloes: theories versus N-body simulations}",
      journal = {\mnras},
         year = 2003,
        month = oct,
       volume = {344},
       number = {4},
        pages = {1327-1333},
          doi = {10.1046/j.1365-8711.2003.06924.x},
archivePrefix = {arXiv},
       eprint = {astro-ph/0306385},
 primaryClass = {astro-ph},
       adsurl = {https://ui.adsabs.harvard.edu/abs/2003MNRAS.344.1327L}
}

@ARTICLE{2018MNRAS.480.5113M,
       author = {{Marinacci}, Federico and {Vogelsberger}, Mark and {Pakmor}, R{\"u}diger and {Torrey}, Paul and {Springel}, Volker and {Hernquist}, Lars and {Nelson}, Dylan and {Weinberger}, Rainer and {Pillepich}, Annalisa and {Naiman}, Jill and {Genel}, Shy},
        title = "{First results from the IllustrisTNG simulations: radio haloes and magnetic fields}",
      journal = {\mnras},
         year = 2018,
        month = nov,
       volume = {480},
       number = {4},
        pages = {5113-5139},
          doi = {10.1093/mnras/sty2206},
archivePrefix = {arXiv},
       eprint = {1707.03396},
 primaryClass = {astro-ph.CO},
       adsurl = {https://ui.adsabs.harvard.edu/abs/2018MNRAS.480.5113M}
}

@ARTICLE{2018MNRAS.477.1206N,
       author = {{Naiman}, Jill P. and {Pillepich}, Annalisa and {Springel}, Volker and {Ramirez-Ruiz}, Enrico and {Torrey}, Paul and {Vogelsberger}, Mark and {Pakmor}, R{\"u}diger and {Nelson}, Dylan and {Marinacci}, Federico and {Hernquist}, Lars and {Weinberger}, Rainer and {Genel}, Shy},
        title = "{First results from the IllustrisTNG simulations: a tale of two elements - chemical evolution of magnesium and europium}",
      journal = {\mnras},
         year = 2018,
        month = jun,
       volume = {477},
       number = {1},
        pages = {1206-1224},
          doi = {10.1093/mnras/sty618},
archivePrefix = {arXiv},
       eprint = {1707.03401},
 primaryClass = {astro-ph.GA},
       adsurl = {https://ui.adsabs.harvard.edu/abs/2018MNRAS.477.1206N}
}

@ARTICLE{2018MNRAS.475..624N,
       author = {{Nelson}, Dylan and {Pillepich}, Annalisa and {Springel}, Volker and {Weinberger}, Rainer and {Hernquist}, Lars and {Pakmor}, R{\"u}diger and {Genel}, Shy and {Torrey}, Paul and {Vogelsberger}, Mark and {Kauffmann}, Guinevere and {Marinacci}, Federico and {Naiman}, Jill},
        title = "{First results from the IllustrisTNG simulations: the galaxy colour bimodality}",
      journal = {\mnras},
         year = 2018,
        month = mar,
       volume = {475},
       number = {1},
        pages = {624-647},
          doi = {10.1093/mnras/stx3040},
archivePrefix = {arXiv},
       eprint = {1707.03395},
 primaryClass = {astro-ph.GA},
       adsurl = {https://ui.adsabs.harvard.edu/abs/2018MNRAS.475..624N}
}

@ARTICLE{2018MNRAS.475..648P,
       author = {{Pillepich}, Annalisa and {Nelson}, Dylan and {Hernquist}, Lars and {Springel}, Volker and {Pakmor}, R{\"u}diger and {Torrey}, Paul and {Weinberger}, Rainer and {Genel}, Shy and {Naiman}, Jill P. and {Marinacci}, Federico and {Vogelsberger}, Mark},
        title = "{First results from the IllustrisTNG simulations: the stellar mass content of groups and clusters of galaxies}",
      journal = {\mnras},
         year = 2018,
        month = mar,
       volume = {475},
       number = {1},
        pages = {648-675},
          doi = {10.1093/mnras/stx3112},
archivePrefix = {arXiv},
       eprint = {1707.03406},
 primaryClass = {astro-ph.GA},
       adsurl = {https://ui.adsabs.harvard.edu/abs/2018MNRAS.475..648P}
}

@ARTICLE{2018MNRAS.475..676S,
       author = {{Springel}, Volker and {Pakmor}, R{\"u}diger and {Pillepich}, Annalisa and {Weinberger}, Rainer and {Nelson}, Dylan and {Hernquist}, Lars and {Vogelsberger}, Mark and {Genel}, Shy and {Torrey}, Paul and {Marinacci}, Federico and {Naiman}, Jill},
        title = "{First results from the IllustrisTNG simulations: matter and galaxy clustering}",
      journal = {\mnras},
         year = 2018,
        month = mar,
       volume = {475},
       number = {1},
        pages = {676-698},
          doi = {10.1093/mnras/stx3304},
archivePrefix = {arXiv},
       eprint = {1707.03397},
 primaryClass = {astro-ph.GA},
       adsurl = {https://ui.adsabs.harvard.edu/abs/2018MNRAS.475..676S}
}

@ARTICLE{1998ApJ...494L...5J,
       author = {{Jing}, Y.~P. and {Suto}, Yasushi},
        title = "{Confronting Cold Dark Matter Cosmologies with Strong Clustering of Lyman Break Galaxies at Z approximately 3}",
      journal = {\apjl},
         year = 1998,
        month = feb,
       volume = {494},
       number = {1},
        pages = {L5-L8},
          doi = {10.1086/311163},
archivePrefix = {arXiv},
       eprint = {astro-ph/9710090},
 primaryClass = {astro-ph},
       adsurl = {https://ui.adsabs.harvard.edu/abs/1998ApJ...494L...5J}
}

@ARTICLE{2002ApJ...574..538J,
       author = {{Jing}, Y.~P. and {Suto}, Yasushi},
        title = "{Triaxial Modeling of Halo Density Profiles with High-Resolution N-Body Simulations}",
      journal = {\apj},
         year = 2002,
        month = aug,
       volume = {574},
       number = {2},
        pages = {538-553},
          doi = {10.1086/341065},
archivePrefix = {arXiv},
       eprint = {astro-ph/0202064},
 primaryClass = {astro-ph},
       adsurl = {https://ui.adsabs.harvard.edu/abs/2002ApJ...574..538J}
}

@ARTICLE{2009ApJ...705..156H,
       author = {{Heitmann}, Katrin and {Higdon}, David and {White}, Martin and {Habib}, Salman and {Williams}, Brian J. and {Lawrence}, Earl and {Wagner}, Christian},
        title = "{The Coyote Universe. II. Cosmological Models and Precision Emulation of the Nonlinear Matter Power Spectrum}",
      journal = {\apj},
         year = 2009,
        month = nov,
       volume = {705},
       number = {1},
        pages = {156-174},
          doi = {10.1088/0004-637X/705/1/156},
archivePrefix = {arXiv},
       eprint = {0902.0429},
 primaryClass = {astro-ph.CO},
       adsurl = {https://ui.adsabs.harvard.edu/abs/2009ApJ...705..156H}
}

@ARTICLE{2010ApJ...715..104H,
       author = {{Heitmann}, Katrin and {White}, Martin and {Wagner}, Christian and {Habib}, Salman and {Higdon}, David},
        title = "{The Coyote Universe. I. Precision Determination of the Nonlinear Matter Power Spectrum}",
      journal = {\apj},
         year = 2010,
        month = may,
       volume = {715},
       number = {1},
        pages = {104-121},
          doi = {10.1088/0004-637X/715/1/104},
archivePrefix = {arXiv},
       eprint = {0812.1052},
 primaryClass = {astro-ph},
       adsurl = {https://ui.adsabs.harvard.edu/abs/2010ApJ...715..104H}
}

@ARTICLE{2021MNRAS.506.2871S,
       author = {{Springel}, Volker and {Pakmor}, R{\"u}diger and {Zier}, Oliver and {Reinecke}, Martin},
        title = "{Simulating cosmic structure formation with the GADGET-4 code}",
      journal = {\mnras},
         year = 2021,
        month = sep,
       volume = {506},
       number = {2},
        pages = {2871-2949},
          doi = {10.1093/mnras/stab1855},
archivePrefix = {arXiv},
       eprint = {2010.03567},
 primaryClass = {astro-ph.IM},
       adsurl = {https://ui.adsabs.harvard.edu/abs/2021MNRAS.506.2871S}
}

@ARTICLE{2024MNRAS.530.2378S,
       author = {{Schaller}, Matthieu and {Borrow}, Josh and {Draper}, Peter W. and {Ivkovic}, Mladen and {McAlpine}, Stuart and {Vandenbroucke}, Bert and {Bah{\'e}}, Yannick and {Chaikin}, Evgenii and {Chalk}, Aidan B.~G. and {Chan}, Tsang Keung and {Correa}, Camila and {van Daalen}, Marcel and {Elbers}, Willem and {Gonnet}, Pedro and {Hausammann}, Lo{\"\i}c and {Helly}, John and {Hu{\v{s}}ko}, Filip and {Kegerreis}, Jacob A. and {Nobels}, Folkert S.~J. and {Ploeckinger}, Sylvia and {Revaz}, Yves and {Roper}, William J. and {Ruiz-Bonilla}, Sergio and {Sandnes}, Thomas D. and {Uyttenhove}, Yolan and {Willis}, James S. and {Xiang}, Zhen},
        title = "{SWIFT: A modern highly-parallel gravity and smoothed particle hydrodynamics solver for astrophysical and cosmological applications}",
      journal = {\mnras},
         year = 2024,
        month = may,
       volume = {530},
       number = {2},
        pages = {2378-2419},
          doi = {10.1093/mnras/stae922},
archivePrefix = {arXiv},
       eprint = {2305.13380},
 primaryClass = {astro-ph.IM},
       adsurl = {https://ui.adsabs.harvard.edu/abs/2024MNRAS.530.2378S}
}

@ARTICLE{2022MNRAS.509.3441A,
       author = {{Anbajagane}, Dhayaa and {Evrard}, August E. and {Farahi}, Arya},
        title = "{Baryonic imprints on DM haloes: population statistics from dwarf galaxies to galaxy clusters}",
      journal = {\mnras},
         year = 2022,
        month = jan,
       volume = {509},
       number = {3},
        pages = {3441-3461},
          doi = {10.1093/mnras/stab3177},
archivePrefix = {arXiv},
       eprint = {2109.02713},
 primaryClass = {astro-ph.CO},
       adsurl = {https://ui.adsabs.harvard.edu/abs/2022MNRAS.509.3441A}
}

@ARTICLE{2013ApJ...762..109B,
       author = {{Behroozi}, Peter S. and {Wechsler}, Risa H. and {Wu}, Hao-Yi},
        title = "{The ROCKSTAR Phase-space Temporal Halo Finder and the Velocity Offsets of Cluster Cores}",
      journal = {\apj},
         year = 2013,
        month = jan,
       volume = {762},
       number = {2},
          eid = {109},
        pages = {109},
          doi = {10.1088/0004-637X/762/2/109},
archivePrefix = {arXiv},
       eprint = {1110.4372},
 primaryClass = {astro-ph.CO},
       adsurl = {https://ui.adsabs.harvard.edu/abs/2013ApJ...762..109B}
}

@ARTICLE{2013ApJ...763...18B,
       author = {{Behroozi}, Peter S. and {Wechsler}, Risa H. and {Wu}, Hao-Yi and {Busha}, Michael T. and {Klypin}, Anatoly A. and {Primack}, Joel R.},
        title = "{Gravitationally Consistent Halo Catalogs and Merger Trees for Precision Cosmology}",
      journal = {\apj},
         year = 2013,
        month = jan,
       volume = {763},
       number = {1},
          eid = {18},
        pages = {18},
          doi = {10.1088/0004-637X/763/1/18},
archivePrefix = {arXiv},
       eprint = {1110.4370},
 primaryClass = {astro-ph.CO},
       adsurl = {https://ui.adsabs.harvard.edu/abs/2013ApJ...763...18B}
}

@ARTICLE{2000ApJ...538..473L,
       author = {{Lewis}, Antony and {Challinor}, Anthony and {Lasenby}, Anthony},
        title = "{Efficient Computation of Cosmic Microwave Background Anisotropies in Closed Friedmann-Robertson-Walker Models}",
      journal = {\apj},
         year = 2000,
        month = aug,
       volume = {538},
       number = {2},
        pages = {473-476},
          doi = {10.1086/309179},
archivePrefix = {arXiv},
       eprint = {astro-ph/9911177},
 primaryClass = {astro-ph},
       adsurl = {https://ui.adsabs.harvard.edu/abs/2000ApJ...538..473L}
}

@ARTICLE{2007ApJ...671.1160L,
       author = {{Luki{\'c}}, Zarija and {Heitmann}, Katrin and {Habib}, Salman and {Bashinsky}, Sergei and {Ricker}, Paul M.},
        title = "{The Halo Mass Function: High-Redshift Evolution and Universality}",
      journal = {\apj},
         year = 2007,
        month = dec,
       volume = {671},
       number = {2},
        pages = {1160-1181},
          doi = {10.1086/523083},
archivePrefix = {arXiv},
       eprint = {astro-ph/0702360},
 primaryClass = {astro-ph},
       adsurl = {https://ui.adsabs.harvard.edu/abs/2007ApJ...671.1160L}
}

@ARTICLE{2018ApJ...859...55C,
       author = {{Child}, Hillary L. and {Habib}, Salman and {Heitmann}, Katrin and {Frontiere}, Nicholas and {Finkel}, Hal and {Pope}, Adrian and {Morozov}, Vitali},
        title = "{Halo Profiles and the Concentration-Mass Relation for a {\ensuremath{\Lambda}}CDM Universe}",
      journal = {\apj},
         year = 2018,
        month = may,
       volume = {859},
       number = {1},
          eid = {55},
        pages = {55},
          doi = {10.3847/1538-4357/aabf95},
archivePrefix = {arXiv},
       eprint = {1804.10199},
 primaryClass = {astro-ph.CO},
       adsurl = {https://ui.adsabs.harvard.edu/abs/2018ApJ...859...55C}
}

@ARTICLE{2012MNRAS.423.3018P,
       author = {{Prada}, Francisco and {Klypin}, Anatoly A. and {Cuesta}, Antonio J. and {Betancort-Rijo}, Juan E. and {Primack}, Joel},
        title = "{Halo concentrations in the standard {\ensuremath{\Lambda}} cold dark matter cosmology}",
      journal = {\mnras},
         year = 2012,
        month = jul,
       volume = {423},
       number = {4},
        pages = {3018-3030},
          doi = {10.1111/j.1365-2966.2012.21007.x},
archivePrefix = {arXiv},
       eprint = {1104.5130},
 primaryClass = {astro-ph.CO},
       adsurl = {https://ui.adsabs.harvard.edu/abs/2012MNRAS.423.3018P}
}

@ARTICLE{2021MNRAS.506.4210I,
       author = {{Ishiyama}, Tomoaki and {Prada}, Francisco and {Klypin}, Anatoly A. and {Sinha}, Manodeep and {Metcalf}, R. Benton and {Jullo}, Eric and {Altieri}, Bruno and {Cora}, Sof{\'\i}a A. and {Croton}, Darren and {de la Torre}, Sylvain and {Mill{\'a}n-Calero}, David E. and {Oogi}, Taira and {Ruedas}, Jos{\'e} and {Vega-Mart{\'\i}nez}, Cristian A.},
        title = "{The Uchuu simulations: Data Release 1 and dark matter halo concentrations}",
      journal = {\mnras},
         year = 2021,
        month = sep,
       volume = {506},
       number = {3},
        pages = {4210-4231},
          doi = {10.1093/mnras/stab1755},
archivePrefix = {arXiv},
       eprint = {2007.14720},
 primaryClass = {astro-ph.CO},
       adsurl = {https://ui.adsabs.harvard.edu/abs/2021MNRAS.506.4210I}
}

@ARTICLE{2013MNRAS.433.1573S,
       author = {{Schneider}, Aurel and {Smith}, Robert E. and {Reed}, Darren},
        title = "{Halo mass function and the free streaming scale}",
      journal = {\mnras},
         year = 2013,
        month = aug,
       volume = {433},
       number = {2},
        pages = {1573-1587},
          doi = {10.1093/mnras/stt829},
archivePrefix = {arXiv},
       eprint = {1303.0839},
 primaryClass = {astro-ph.CO},
       adsurl = {https://ui.adsabs.harvard.edu/abs/2013MNRAS.433.1573S}
}

@ARTICLE{2001ApJ...559..516A,
       author = {{Avila-Reese}, Vladimir and {Col{\'\i}n}, Pedro and {Valenzuela}, Octavio and {D'Onghia}, Elena and {Firmani}, Claudio},
        title = "{Formation and Structure of Halos in a Warm Dark Matter Cosmology}",
      journal = {\apj},
         year = 2001,
        month = oct,
       volume = {559},
       number = {2},
        pages = {516-530},
          doi = {10.1086/322411},
archivePrefix = {arXiv},
       eprint = {astro-ph/0010525},
 primaryClass = {astro-ph},
       adsurl = {https://ui.adsabs.harvard.edu/abs/2001ApJ...559..516A}
}

@ARTICLE{2019MNRAS.488.3663L,
       author = {{Ludlow}, Aaron D. and {Schaye}, Joop and {Bower}, Richard},
        title = "{Numerical convergence of simulations of galaxy formation: the abundance and internal structure of cold dark matter haloes}",
      journal = {\mnras},
         year = 2019,
        month = sep,
       volume = {488},
       number = {3},
        pages = {3663-3684},
          doi = {10.1093/mnras/stz1821},
archivePrefix = {arXiv},
       eprint = {1812.05777},
 primaryClass = {astro-ph.CO},
       adsurl = {https://ui.adsabs.harvard.edu/abs/2019MNRAS.488.3663L}
}

@ARTICLE{2010MNRAS.402...21N,
       author = {{Navarro}, Julio F. and {Ludlow}, Aaron and {Springel}, Volker and {Wang}, Jie and {Vogelsberger}, Mark and {White}, Simon D.~M. and {Jenkins}, Adrian and {Frenk}, Carlos S. and {Helmi}, Amina},
        title = "{The diversity and similarity of simulated cold dark matter haloes}",
      journal = {\mnras},
         year = 2010,
        month = feb,
       volume = {402},
       number = {1},
        pages = {21-34},
          doi = {10.1111/j.1365-2966.2009.15878.x},
archivePrefix = {arXiv},
       eprint = {0810.1522},
 primaryClass = {astro-ph},
       adsurl = {https://ui.adsabs.harvard.edu/abs/2010MNRAS.402...21N}
}

@ARTICLE{2003MNRAS.338...14P,
       author = {{Power}, C. and {Navarro}, J.~F. and {Jenkins}, A. and {Frenk}, C.~S. and {White}, S.~D.~M. and {Springel}, V. and {Stadel}, J. and {Quinn}, T.},
        title = "{The inner structure of {\ensuremath{\Lambda}}CDM haloes - I. A numerical convergence study}",
      journal = {\mnras},
         year = 2003,
        month = jan,
       volume = {338},
       number = {1},
        pages = {14-34},
          doi = {10.1046/j.1365-8711.2003.05925.x},
archivePrefix = {arXiv},
       eprint = {astro-ph/0201544},
 primaryClass = {astro-ph},
       adsurl = {https://ui.adsabs.harvard.edu/abs/2003MNRAS.338...14P}
}
\bibliographystyle{aasjournalv7}

\end{document}